\documentclass[traditabstract]{aa} 
\usepackage{graphicx}
\usepackage{marginnote}
\usepackage{txfonts}
\usepackage{subfigure}
\usepackage{natbib}
\usepackage{amstext}
\usepackage[verbose]{placeins}
\usepackage{tikz}
\usepackage{array}
\usepackage{amssymb}
\usepackage{xcolor}

\newcommand{\bd}{\begin{displaymath}}
\newcommand{\ed}{\end{displaymath}}
\newcommand{\be}{\begin{equation}}
\newcommand{\ee}{\end{equation}}
\newcommand{\beaa}{\begin{eqnarray*}}
\newcommand{\eeaa}{\end{eqnarray*}}
\newcommand{\bea}{\begin{eqnarray}}
\newcommand{\eea}{\end{eqnarray}}

\begin{document}

   \title{HOLISMOKES - IV. Efficient mass modeling of strong lenses through deep learning}

   \titlerunning{HOLISMOKES - IV. Efficient mass modeling of strong lenses through deep learning}

   \author{S.~Schuldt\inst{1}\inst{,2}
          \and
          S.~H.~Suyu\inst{1}\inst{,2}\inst{,3}
          \and
          T.~Meinhardt\inst{4}
          \and
          L.~Leal-Taix\'{e}\inst{4}
          \and
          R.~Ca\~{n}ameras\inst{1}
          \and
          S.~Taubenberger\inst{1}
          \and
          A.~Halkola\inst{5}
          }

   \institute{Max-Planck-Institut f\"ur Astrophysik, Karl-Schwarzschild Str.~1, 85741 Garching, Germany \\
              \email{schuldt@mpa-garching.mpg.de}
         \and  
             Physik Department, Technische Universit\"at M\"unchen, James-Franck Str. 1, 85741 Garching, Germany
         \and
             Institute of Astronomy and Astrophysics, Academia Sinica, 11F of ASMAB, No.1, Section 4, Roosevelt Road, Taipei 10617, Taiwan
         \and
            Informatik Department, Technische Universit\"at M\"unchen, Bolzmannstr. 3, 85741 Garching, Germany
         \and
            Py\"orrekuja 5 A, FI-04300 Tuusula, Finland
             }

   \date{Received --; accepted --}


  \abstract
      {
      Modeling the mass distributions of strong gravitational lenses
      is often necessary in order to use them as astrophysical and cosmological
      probes.
      With the large number of lens systems ($\gtrsim$$10^5$)
      expected from upcoming surveys, it is timely to explore
      efficient modeling approaches beyond traditional Markov chain
      Monte Carlo techniques that are time consuming.  We train a
      convolutional neural network (CNN) on images of galaxy-scale
      lens systems to predict the five parameters of the singular
      isothermal ellipsoid (SIE) mass model (lens center $x$ and $y$,
      complex ellipticity $e_x$ and $e_y$, and Einstein radius
      $\theta_\text{E}$). To train the network we simulate images
      based on real observations from the Hyper Suprime-Cam Survey for
      the lens galaxies and from the Hubble Ultra Deep Field as lensed
      galaxies.  We tested different network architectures and the
      effect of different data sets, such as using only double or quad
      systems defined based on the source center and using different input distributions of
      $\theta_\text{E}$. We find that the CNN performs well, and 
      with the network trained on both doubles and quads with a
      uniform distribution of $\theta_\text{E}$ $>0.5\arcsec$ we obtain the
      following median values with $1\sigma$ scatter:
      $\Delta x = (0.00^{+0.30}_{-0.30}) \arcsec$,
      $\Delta y = (0.00^{+0.30}_{-0.29} ) \arcsec $,
      $\Delta \theta_\text{E} = (0.07^{+0.29}_{-0.12}) \arcsec$,
      $\Delta e_x = -0.01^{+0.08}_{-0.09}$, and
      $\Delta e_y = 0.00^{+0.08}_{-0.09}$. 
      The bias in
      $\theta_\text{E}$ is driven by systems with small
      $\theta_\text{E}$. Therefore, when we further predict the
      multiple lensed image positions and \mbox{time-delays} based on the
      network output, we apply the network to the sample limited to
      $\theta_\text{E}>0.8\arcsec$. In this case the offset between
      the predicted and input lensed image positions is
      $(0.00_{-0.29}^{+0.29}) \arcsec$ and
      $(0.00_{-0.31}^{+0.32}) \arcsec$ for the $x$ and $y$ coordinates,
      respectively. For the fractional difference between the
      predicted and true \mbox{time-delay}, we obtain
      $0.04_{-0.05}^{+0.27}$. Our CNN model is able to predict the
      SIE parameter values in fractions of a second on a single CPU,
      and with the output we can predict the image positions and \mbox{time-delays} in an automated way, such that we are able to process
      efficiently the huge amount of expected galaxy-scale lens
      detections in the near future.}


   \maketitle
%
\section{Introduction}
\label{sec:introduction}

Strong gravitational lensing has become a very powerful tool for probing 
various properties of the Universe. For instance, galaxy-galaxy
lensing can help to constrain the total mass of the lens and, assuming a mass-to-light ratio (M/L)  for the baryonic
matter, also its dark matter (DM) fraction. By combining lensing with
other methods like measurements of the lens' velocity dispersion
\citep[e.g.,][]{barnabe11, barnabe12, yildirim20} or the galaxy rotation curves
\citep[e.g.,][]{hashim14, strigari13}, the dark matter can be better
disentangled from the baryonic component and a 3D (deprojected) model
of the mass density profile can be
obtained. Such profiles are very helpful for probing cosmological
models \citep[e.g.,][]{davies18, eales15, krywult17}.

Another application of strong lensing is to probe \mbox{high-redshift}
sources thanks to the lensing magnification \citep[e.g.,][]{dye18, lemon18,
  mcgreer18, rubin18, salmon18, shu18}. In recent years, huge efforts
have been made  in reconstructing the surface brightness distribution of
lensed extended sources. Together with redshift and kinematic measurements, these
observations contain information about the evolution of galaxies at higher
reshifts. If the mass profile of the lens is well constrained, the
original unlensed morphology can be reconstructed 
\citep[e.g.,][]{warren03, suyu06, nightingale18, rizzo18, chirivi20}.

Lensed supernovae (SNe) and lensed quasars are very powerful
cosmological probes. By measuring the \mbox{time-delays} of a lensing system
with an object that is variable in brightness, one can  use it to
constrain, for example, the Hubble constant $H_0$ \citep[e.g.,][]{refsdal64, chen19,
  rusu20, wong20, shajib20}. This helps to assess the $4.4 \sigma$ tension between the cosmic microwave background
(CMB) analysis that gives $H_0 = (67.36 \pm 0.54)~ \text{km s}^{-1} \text{Mpc}^{-1}
$ for flat $\Lambda$ cold dark matter
\citep[$\Lambda$CDM;][]{PlankCollaboration18} and the local distance
ladder with $H_0 = (74.03 \pm 1.42)~ \text{km s}^{-1} \text{Mpc}^{-1}
$ \citep[SH0ES project;][]{riess19}. To date, \mbox{time-delay} lensing cosmography has   been mainly
based on lensed quasars as the chance of a lensed supernova (SN) is substantially lower.  There are currently two lensed SNe known: one \mbox{core-collapse} SN behind a
strong lensing cluster MACS J1149.5+222.3 \citep[SN
Refsdal;][]{kelly15} and one SN type Ia behind an isolated lens galaxy
\citep[iPTF16geu;][]{goobar17}. Thanks to the upcoming wide field surveys in the
next decades, like the Rubin Observatory Legacy Survey of Space and Time \citep[LSST,][]{ivezic08}, this will change. LSST is expected to detect
hundreds of lensed SNe \citep[e.g.,][]{goldstein19, wojtak19}.
Therefore, it
is important to be prepared for such exciting transient events in a
fully automated and fast way. In particular, a fast
  estimation of \mbox{time-delay(s)} is important for optimizing the
  observing--monitoring strategy for \mbox{time-delay} measurements.

In addition to  \mbox{time-delay} measurements, observing lensed SNe type Ia can help
to answer outstanding questions about their progenitor systems \citep{suyu20}. The basic scenario is the single degenerate case where a white dwarf (WD) is stable until it reaches the Chandrasekhar mass limit \citep{whelan73, nomoto82} by accreting mass from a nearby star. Today there are also alternative scenarios considered where the WD explodes before reaching the Chandrasekhar mass, the so-called sub-Chandrasekhar detonations \citep{sim10}. Another possibility for a SN Ia is the \mbox{double-degenerated} scenario where the companion is another WD \citep[e.g.,][]{pakmor10} and both are merging to exceed the Chandrasekhar mass limit. It is still unclear which  of the main scenarios is correct to describe the SN Ia formation, or if both are. To shed light on this debate, one possibility is to observe the SN Ia spectroscopically at very early stages, which is normally difficult because SN detections are often
close to peak luminosity, past the early phase. If this SN is lensed, we can use the
position of the first appearing image, 
together with a mass model of the underlying lens galaxy,
to predict the position and time
when the next images will appear. Here it is very important to react
quickly,
particularly to compute the mass model of the underlying lens galaxy based on imaging,
as the \mbox{time-delays} of galaxy-galaxy strong lensing are typically
on the order of days to weeks.

Since these strong lens observations are very powerful, several large
surveys including the Sloan Lens ACS (SLACS) survey \citep{bolton06, shu17}, the
CFHTLS Strong Lensing Legacy Survey \citep[SL2S;][]{cabanac07, sonnenfeld15}, the
Sloan WFC Edge-on Late-type Lens Survey \citep[SWELLS;][]{treu11}, the
BOSS Emission-Line Lens Survey \citep[BELLS;][]{brownstein12, shu16, cornachione18},
the Dark Energy Survey \citep[DES;][]{DES05, tanoglidis20}, the Survey of
Gravitationally-lensed Objects in HSC Imaging
\citep[SuGOHI;][]{sonnenfeld18, wong18, chan20, jaelani20}, and surveys in the Panoramic Survey Telescope and Rapid Response System \citep[Pan-STARRS; e.g.,][]{lemon18, canameras20} have been conducted
to find lenses. So far we have detected  several thousand lenses,
 but mainly from the lower redshift regime. However, based on newer upcoming surveys like the LSST, which will target around $20,000$ $ \text{deg}^2$ of
the southern hemisphere in six different filters $(u, g, r, i, z,
y)$, together with the Euclid imaging survey from space operated by the European Space
Agency \citep[ESA;][]{laureijs11}, we expect billions of galaxy images containing on
the order of one hundred thousand lenses \citep{collett15}.

To deal with this huge amount of images there are ongoing efforts to
develop fast and automated algorithms to find lenses in the first place. These methods are based on different identification
properties, for instance on geometrical quantification \citep{bom17,
  seidel07}, spectroscopic analysis \citep{baron17, ostrovski17}, or
color cuts \citep{gavazzi14, maturi14}. Moreover,
convolutional neural networks (CNNs) have also been extensively used
in gravitational lens detection \citep[e.g.,][]{jacobs17, petrillo17, schaefer18, lanusse18, metcalf19, canameras20, huang20} 
as they do not require any
measurements of the lens properties. Once a CNN is trained, it can
classify huge amounts of images in a very short time, and is thus very
efficient. Nonetheless,  CNNs have limitations (e.g.,  completeness or  accurate grading) and the performance strongly depends on the training set design as it encodes an effective prior (in the case of supervised learning). In this regard unsupervised or active learning might be promising future avenues for finding lenses.

However, these methods are only for finding the lenses; a mass model is necessary for further studies. Mass models of
gravitational lenses are often described by  parameterized profiles, where
the parameters are optimized, for instance via Markov chain Monte Carlo (MCMC)
sampling \citep[e.g.,][]{jullo07, suyu10, sciortino20, fowlie20}.
 These techniques are very time
and resource consuming as modeling one lens can take weeks or months, and they are thus difficult to scale up for the upcoming
amount of data. With the success of CNNs in image processing,
\citet{hezaveh17} showed the use of CNNs in estimating the mass model
parameters of a singular isothermal ellipsoid (SIE)
profile, and investigated further error estimations \citep{levasseur17}, analysis of interferometric observations \citep{morningstar18}, and source surface brightness reconstruction with recurrent inference machines \citep[RIMs;][]{morningstar19}. While
they  mainly consider \mbox{single-band} images and subtract the lens light
before processing the image with the CNN, \citet{pearson19} presented
a CNN to model the image without lens light subtraction. However, for
all deep learning approaches one needs a data set that contains the
images and the corresponding parameter values for training,
validation, and testing the network. As there are not that many real lensed
galaxies known, both groups use mock lenses for their CNNs.

We recently initiated the Highly Optimized Lensing Investigations of
Supernovae, Microlensing Objects, and Kinematics of Ellipticals and
Spirals (HOLISMOKES) program \citep[][hereafter HOLISMOKES
I]{suyu20}. After presenting our 
lens search project \citep[][hereafter HOLISMOKES II]{canameras20}, we present
in this paper a CNN for modeling strong gravitationally lensed
galaxies with \mbox{ground-based} imaging, 
taking  advantage of four different filters and not applying lens light
subtraction beforehand. In contrast to \citet{pearson19}, we  use
a \mbox{mocked-up} data set based on real observed galaxy cutouts since the
performance of the CNN on real systems will be optimal when the mock
systems used for training are as close to real lens observations as
possible. 
Our mock lens images contain, by construction, realistic \mbox{line-of-sight}
objects as well as realistic lens and source light distributions in the
image cutouts. We
  use the Hyper \mbox{Suprime-Cam} (HSC) Subaru Strategic Program (SSP) images
  together with redshift and velocity dispersion measurements
from the Sloan Digital Sky Survey (SDSS) for the lens galaxies, and  images
together with redshifts from the Hubble Ultra Deep Field (HUDF)
survey for the sources \citep{beckwith06, inami17}.

The outline of the paper is as follows. We describe in
Sect.~\ref{sec:simulation} how we simulate our training data, and we
give a short introduction and overview of the used network
architecture in Sect.~\ref{sec:network}. The main networks are
presented in Sect.~\ref{sec:results}, and we give details of further
tests in Sect. \ref{sec:tests}. We also consider the image position and
\mbox{time-delay} differences in Sect.~\ref{sec:timedelay} for a performance
test, and compare them to other modeling techniques in Sect.~\ref{sec:otherTechniques}. We summarize and conclude our results  in
Sect.~\ref{sec:conclusion}.
Throughout this work we assume a flat $\Lambda$CDM cosmology with a
Hubble constant $H_0 = 72\, \text{km}\, \text{s}^{-1}\,
\text{Mpc}^{-1}$ \citep{bonvin17} and $\Omega_\text{M} =1
-\Omega_\Lambda = 0.32 $ \citep{PlankCollaboration18}. Unless
specified otherwise, each quoted parameter estimate is the median of its
1D marginalized posterior probability density function, 
and the quoted uncertainties show the 16$\text{th}$ and 84$\text{th}$ percentiles
(i.e., the bounds of a 68\% credible interval).

\FloatBarrier
\section{Simulation of strongly lensed images}
\label{sec:simulation}

For training a neural network one needs, depending on the network size,
from tens of thousands   to millions of images, together with the expected network
output, which in our case are the values of the SIE profile parameters
corresponding to each image. Since there are far too few known lens
systems, we need to mock up lens images. While previous studies are
based on partly or fully generated light distributions \citep[e.g.,][]{hezaveh17,
  levasseur17, pearson19}, we aim to produce more realistic
lens images by using real
observed images of galaxies and  simulating only the lensing effect
with our own routine. We work with the four HSC filters, $g, r, i$, and $z$ 
(respectively matched to HST filters F435W ($\overline{\lambda} = 4343.4 \AA$), 
F606W ($\overline{\lambda} =
6000.8 \AA$), F775W ($\overline{\lambda} = 7702.2 \AA$), and
F850LP ($\overline{\lambda} = 9194.4 \AA$)) to give the network
the color information to distinguish better between lens and source galaxies. The images of HSC for these filters are very similar to the expected image quality of LSST, such that our tests and findings will also hold for LSST. Therefore, this work is in direct preparation for and an important step in modeling the expected 100,000 lens systems that will be detected with LSST in the near future.

\subsection{Lens galaxies from HSC}
\label{sec:simulation:lens}

\begin{figure}[ht!]
  \includegraphics[trim=0 0 0 0, clip, width=0.47\textwidth]{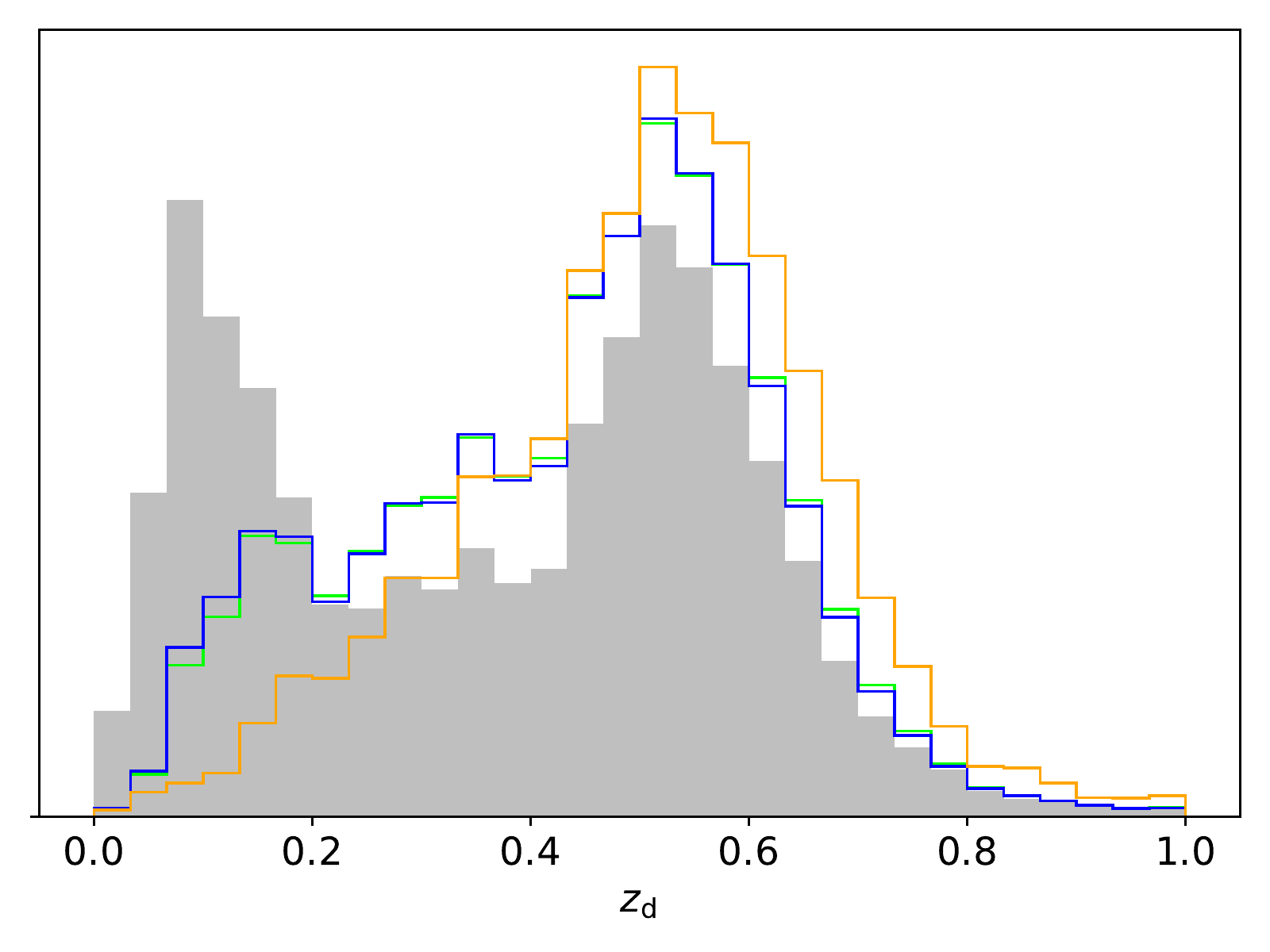}
  \includegraphics[trim=0 0 0 0, clip, width=0.47\textwidth]{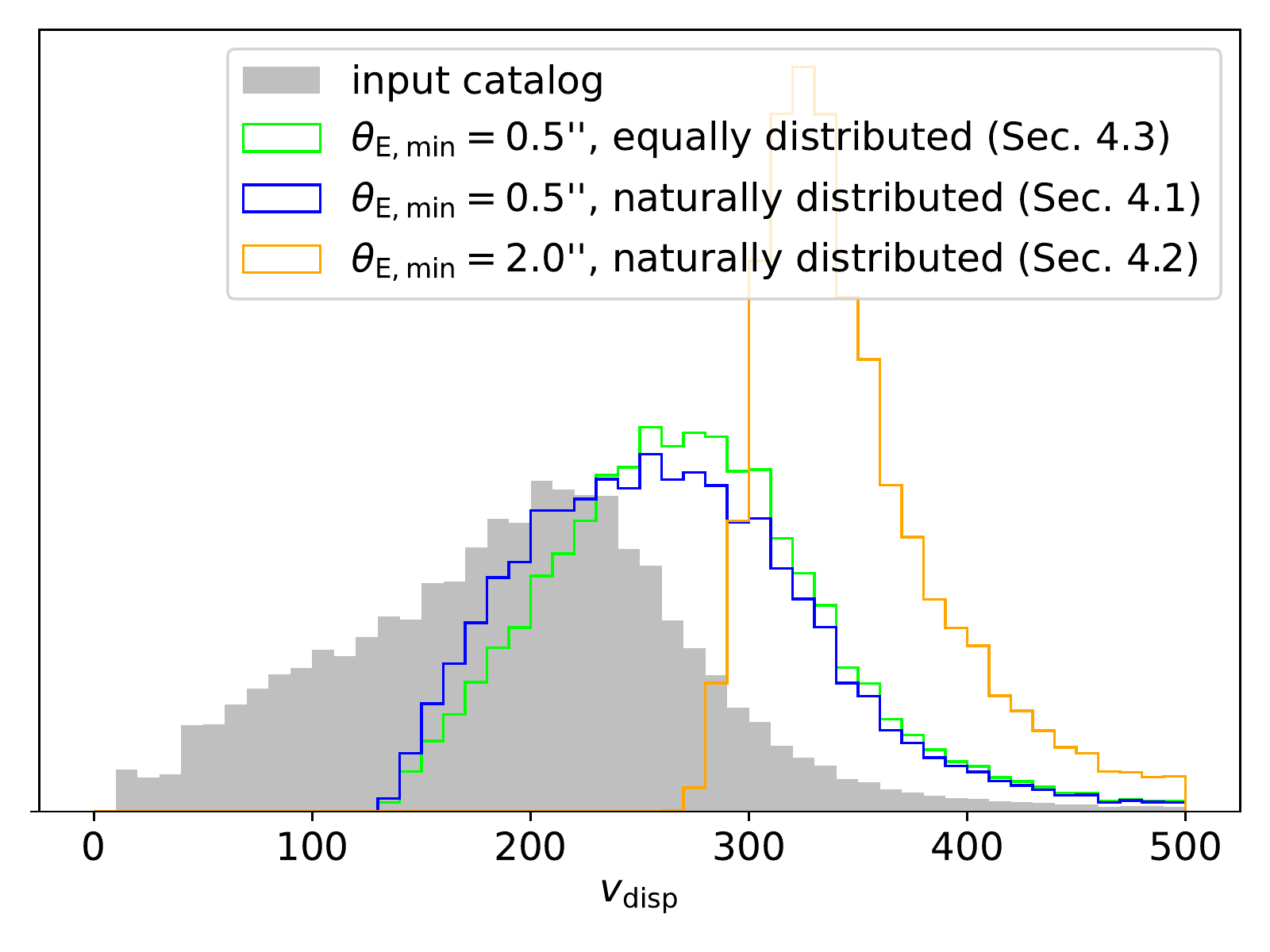}
\caption{Distributions of the lens galaxy redshifts $z_\text{d}$ (top) and velocity dispersion $v_\text{disp}$ (bottom). Shown are the distributions of the input catalogs to the simulation code (in gray), and the distributions of the generated samples discussed in Sect. \ref{sec:results} (see inset for color-coding).}
\label{fig:hist_lens}
\end{figure}

For the lenses we use HSC SSP\footnote{HSC SSP webpage:
  {https://hsc-release.mtk.nao.ac.jp/doc/}} images from the second
public data release \citep[PDR2;][]{aihara19} with a pixel size of $0.168\arcsec$. To calculate the axis
ratio $q_\text{light}$ and position angle $\theta_\text{light} $ 
 of the lens, we use the second brightness  moments calculated for the $i$
band since redder filters follow better the stellar mass; however,  the
S/N is substantially lower in the $z$ band compared to the $i$
band. We cross-match the HSC catalog with the 
SDSS\footnote{SDSS webpage: https://www.sdss.org/; catalog downloaded
  from the 14$\text{th}$ data release page
  http://skyserver.sdss.org/dr14/en/tools /search/sql.aspx .} catalog
to use only images of galaxies where we have SDSS spectroscopic redshifts and velocity dispersions. With this selection  we end up with a sample containing 145,170 galaxies that is dominated by luminous red galaxies (LRGs). We show in Figure~\ref{fig:hist_lens}   a histogram of the lens redshifts used for the simulation (in gray). We already overplot the distribution of the mock samples discussed in Sect.~\ref{sec:results}.

To describe the mass distribution of the lens, we
adopt a SIE profile \citep{ 
  barkana98} such that the convergence (dimensionless surface
mass density) can be expressed as 
\be
\kappa(x,y) = \frac{\theta_{\text{E}}}{(1+q) r} 
\ee
with elliptical radius
\be
r =\sqrt{x^2+\frac{y^2}{q^2}},
\ee
where $x$ and $y$ are angular coordinates on the lens plane with respect to the lens center.
In this equation
$\theta_\text{E}$ denotes the Einstein radius and $q$ the axis ratio.\footnote{The SIE mass profile introduced by \citet{barkana98} allows for an additional core radius, which we set to $10^{-4}$, that yields effectively a singular mass distribution without numerical issues at the lens center.}
The mass distribution is rotated by the position angle $\theta$.
The Einstein radius is obtained from the velocity
dispersion $v_\text{disp}$ with
\be
\theta_\text{E} =  4\pi \frac{v_\text{disp}^2}{c^2} \frac{D_\text{ds}}{D_\text{s}} ,
\label{eq:EinsteinRadius}
\ee
where $c$ is the speed of light, and $D_\text{ds}$ and $D_\text{s}$
are respectively the angular diameter distances between the lens (deflector) and source and
the observer and source. The distribution of the velocity dispersion is shown in Figure~\ref{fig:hist_lens} (bottom pannel, gray histogram). We compute the deflection angles of the
SIE with the lensing software {\sc Glee} \citep{suyu10, suyu12}.

Based on
the second brightness moments of the lens light distribution in the $i$
band, the axis ratio $q_\text{light}$ and position angle
$\theta_\text{light}$ are obtained internally in our simulation code. Based on several
studies \citep[e.g,][]{sonnenfeld18b, loubser20}, the light traces the
mass relatively well but not 
perfectly. Therefore, we add randomly drawn Gaussian perturbations on the light
parameters, with a Gaussian width of $0.05\arcsec$ for the lens
center,   $0.05$ for the axis ratio, and $0.17$ radians (10 degrees) for the
position angle, and adopt the resulting parameter values for the lens
mass distribution. If the axis ratio of the mass $q$ (i.e., with
Gaussian perturbation) is above 1, we draw a second realization of the
Gaussian noise. If the resulting $q$ (from the second Gaussian perturbation) is $\leq1$, then we keep this value; otherwise, we set $q$ to exactly 1.

While the simulation code assumes a parameterization in terms of axis ratio $q$ and position angle $\theta$, we parameterize for our network in terms of complex ellipticity $e_\text{c}$, which we define as $e_\text{c} = A ~e^{2i\theta} = e_x +i e_y$ with
\begin{eqnarray}
e_x&=& \frac{1-q^2}{1+q^2} \, \cos(2\theta), \nonumber \\
e_y&=& \frac{1-q^2}{1+q^2} \, \sin(2\theta).
\end{eqnarray}
The back transformation is given by
\begin{equation}
  \begin{array}{ll}
  q = & \sqrt{\frac{1-\sqrt{e_x^2+e_y^2}}{1+\sqrt{e_x^2+e_y^2}}} \label{eq:backtrafo}\\
  \theta = &
  \left\{
    \begin{array}{ll}
      \frac{1}{2} \arccos \left(e_y \, \frac{1+q^2}{1-q^2} \right) & \text{ if } e_x> 0\\
      \frac{\pi}{2} + \left| \frac{1}{2} \arcsin \left(e_x \, \frac{1+q^2}{1-q^2} \right) \right| & \text{ if } e_x< 0
    \end{array}.
    \right.
  \end{array}
\end{equation}
This is in agreement with previous CNN applications to lens modeling \citep{hezaveh17, pearson19}.

\subsection{Sources from HUDF}
\label{sec:simulation:source}

\begin{figure}[ht!]
\includegraphics[trim=0 0 0 0, clip, width=0.47\textwidth]{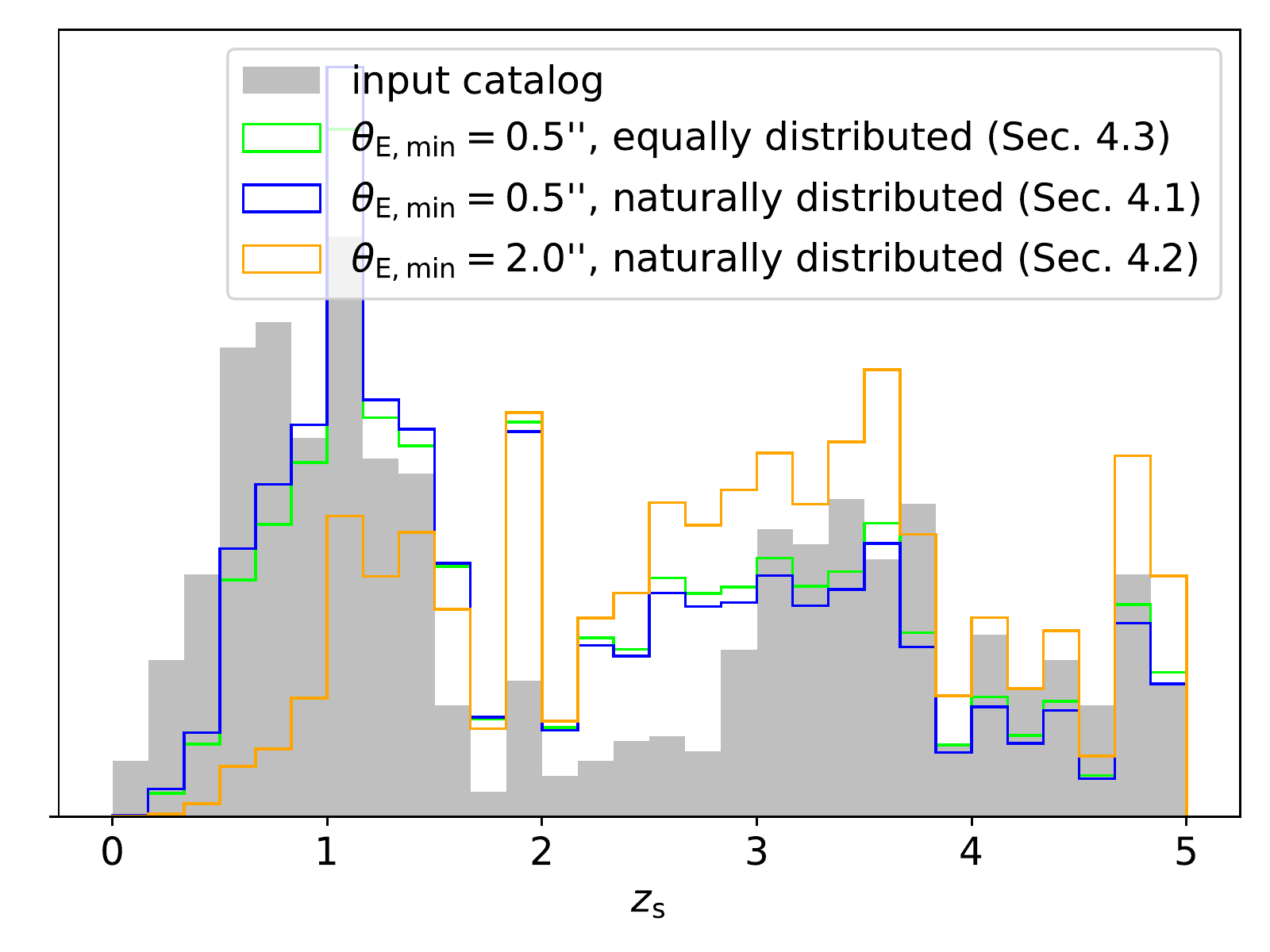}
\caption{Distributions of the source redshifts $z_\text{s}$ of the input catalog to the simulation code (gray) and of the different mock samples (colors) discussed in Sect.~\ref{sec:results}.}
\label{fig:hist_zs}
\end{figure}

The images for the sources are taken from HUDF\footnote{HUDF webpage
  https://www.spacetelescope.org/; downloaded on Oct. 1, 2018, from
  https://archive.stsci.edu/pub/hlsp/udf/acs-wfc/.} where  the spectroscopic
redshifts are also known \citep{beckwith06, inami17}. The cutouts are approximately $10\arcsec
\times 10\arcsec$ with a pixel size of $0.03\arcsec$. This
survey is chosen for its high spatial resolution, and we can adopt
the images without point spread function (PSF) deconvolution. Moreover, it contains \mbox{high-redshift} galaxies such that we can achieve a realistic lensing effect. The 1,323 relevant galaxies are extracted with Source Extractor \citep{bertin96}
since the lensing effect is redshift dependent and we would otherwise
lens the neighboring objects as if they were all at the same
redshift, which would lead to incorrect lensing features. We show a histogram of the source redshifts in Figure~\ref{fig:hist_zs} (gray histogram). Since we randomly select a background source (see Sect.~\ref{sec:simulation:mock} for details), the source galaxies can be used multiple times for one mock sample, and thus the redshift distribution varies slightly between the different samples (colored histograms; see details in Sect.~\ref{sec:results}).

\subsection{Mock lens systems}
\label{sec:simulation:mock}

\begin{figure}[ht!]
\includegraphics[trim=30 0 470 0, clip, width=0.5\textwidth]{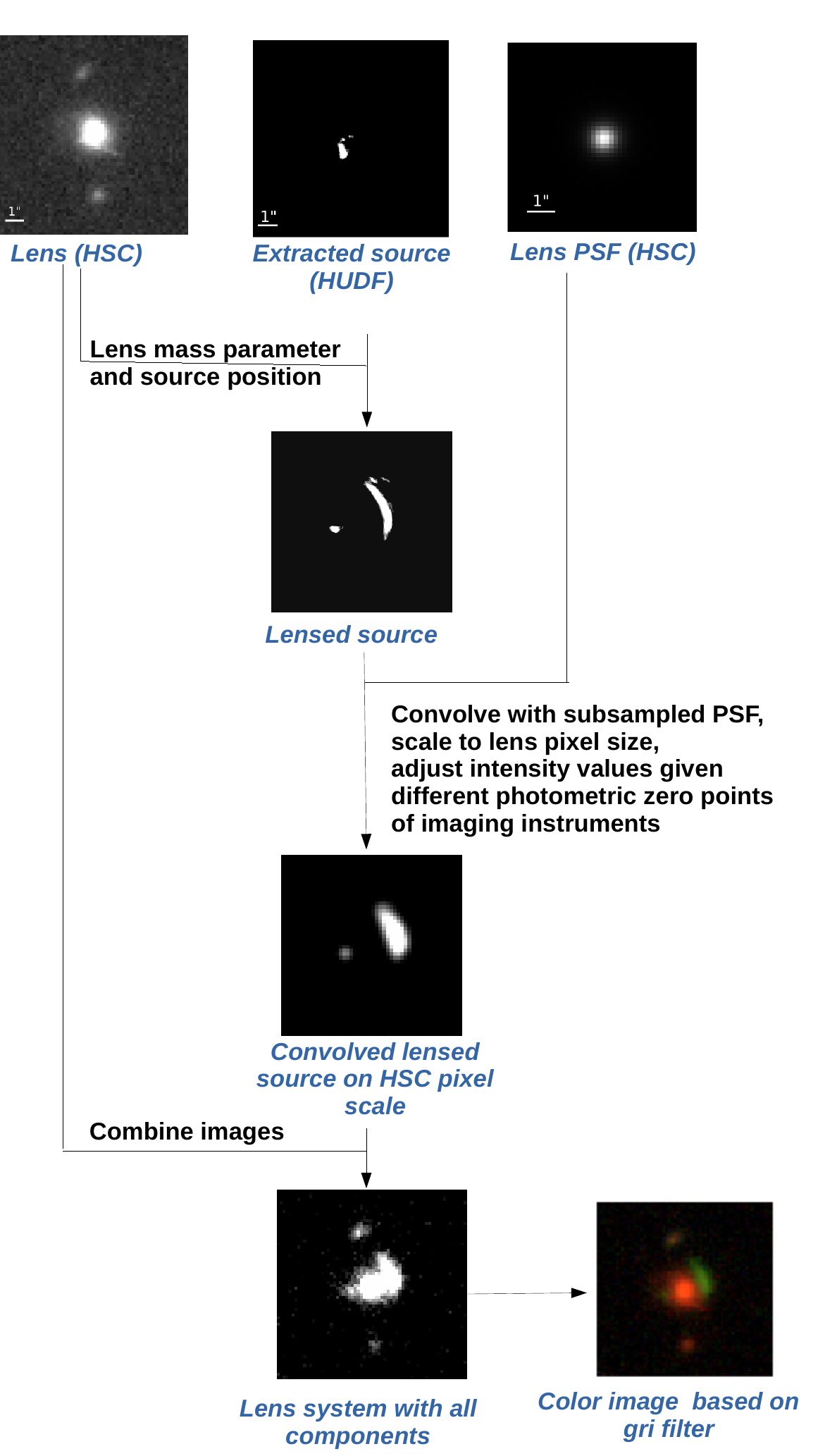}
\caption{Diagram of the simulation pipeline.}
\label{fig:simulation}
\end{figure}

To train our networks we use mock images based on real
observed galaxies, and only generate the lensing effect. We use HSC
galaxies as lenses (see Sect. \ref{sec:simulation:lens} for details)
and HUDF galaxies as background objects (see
Sect. \ref{sec:simulation:source}) to obtain mocks that are as realistic as possible. Figure \ref{fig:simulation} shows a
diagram of the simulation pipeline. The input has three images:
the lens, the unlensed source, and the lens PSF image (top row). Together with the provided redshifts of source and
lens, as well as the velocity dispersion for calculating the Einstein radius with equation
\ref{eq:EinsteinRadius}, the source image can be
lensed onto the lens plane (second row). For this we place a random
source from our catalog randomly in a specified region behind the lens
and accept this position if we obtain a strongly lensed image. Since
the source images have previously been extracted, we use the brightest pixel
in the $i$ band to center the source. We have also implemented the
option to just keep one of the two strong lens configurations, either
quadruply or doubly imaged galaxies, classified based on the image multiplicity of the lensed source
center.  
We also set a peak brightness
threshold for the arcs based on the background noise of the lens. To estimate the background noise we take the corner with size 10\%$\times$10\% square of the full lens image (rounded to an integer of pixel) with the lowest maximum and compute from the patch the root mean square (RMS) value  used as background noise. We take the lowest maximum for each corner
separately and then compute the RMS of that one  because there might be
line-of-sight objects in the corners that would raise the RMS values. To avoid contamination to the background estimation from the lens, we use $40\arcsec \times 40\arcsec$ image cutouts such that each corner is $4\arcsec \times 4\arcsec$. The peak brightness of the lensed source must then be higher than the RMS to be accepted by the simulation code.

\begin{figure}[ht!]
\includegraphics[trim=0 0 0 0, clip, width=0.5\textwidth]{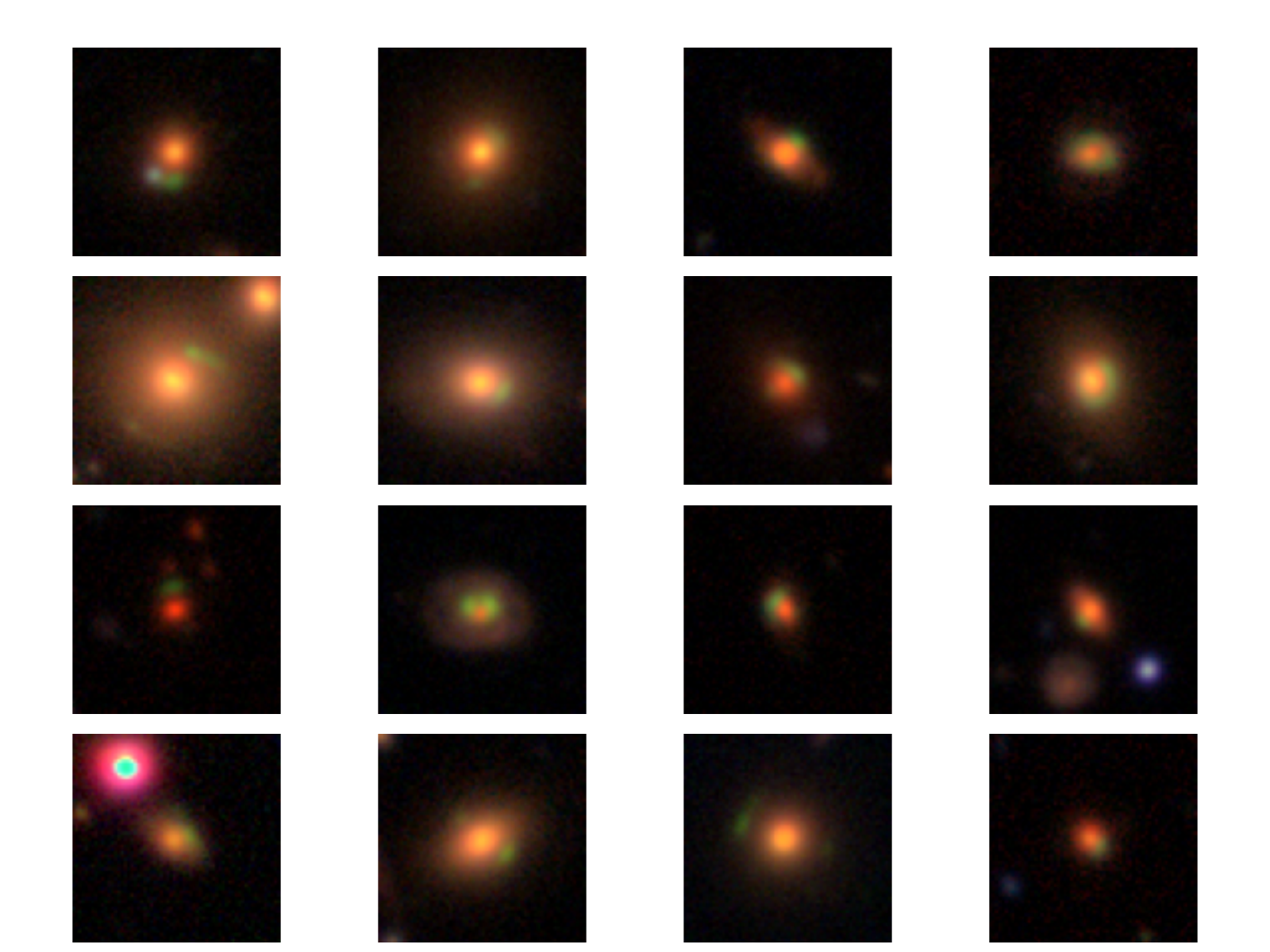}
\caption{Examples of strong gravitational lens systems mocked up with our simulation code by using HUDF galaxies as sources behind HSC galaxies as lenses.  Each image cutout is $10.8\arcsec\times10.8\arcsec$.}
\label{fig:mosaic}
\end{figure}

In the next step the lensed source image with high resolution is convolved
with the subsampled PSF of the lens, which is
provided by HSC SSP PDR2 for each image separately. After binning  the \mbox{high-resolution} lensed, convolved source image
to the HSC pixel size and accounting for the different photometric zeropoints of the source telescope $\text{zp}_\text{sr}$ and lens telescope $\text{zp}_\text{ls}$, which gives a factor of $10^{0.4(\text{zp}_\text{ls}-\text{zp}_\text{sr})}$, the lensed source image is
obtained as if it had been observed through the HSC instrument (third
row in Figure \ref{fig:simulation}), i.e., on the HSC 0.168\arcsec/pixel
resolution.  At this point we neglect the additional Poisson noise for the lensed arcs. Finally, the original lens and the mock lensed source images can
be combined, which results in the final image (fourth row) that is cropped to a size of $64 \times 64$ pixels ($10.8 \arcsec \times 10.8 \arcsec$). For better
illustration, a color image based on the filters $g$,
$r$, and $i$ is also shown, but we generate all mock images in four
bands, which we use for the network training. We show more example images based on $gri$ filters in Figure~\ref{fig:mosaic}. During this simulation, we set an upper limit on the Einstein radius of $5\arcsec$, which corresponds to the size of the biggest Einstein radius so far observed from galaxy-galaxy lensing
\citep{belokurov07}.

We test the effect of different assumptions on the data set, like splitting up in
quads-only or doubles-only, or different assumptions on the
distribution of the Einstein radii since we found this to be crucial
for the network performance. For this we generate with this pipeline new independent mock images that are based on the same lens and source images, but different combinations and alignments. The details of the different samples and the network trained on them will be discussed further in Sect.~\ref{sec:results}. For the set of quads-only and higher limit on the
Einstein radius of 2\arcsec\ we 
use a modification of the conventional data augmentation in deep learning.
In particular, we rotate only the lens image before adding the random lensed
source image, but not the whole final image (which is done normally
for data augmentation). Thus, the ground truth values are also not exactly the same values given the change in position angle and another background source with different location and redshift.

\section{Neural networks and their architecture}
\label{sec:network}

Neural networks (NNs) are extremely powerful tools for a wide range of 
tasks, and thus in  recent years broadly used and
explored. Additionally, the computational time can be reduced notably compared to other methods.
There are generally
two types of NNs: (1) classification, where the  ground truth uses different labels to distinguish between the different classes, and  (2) regression, where the ground truth consists of a set of parameters with specific values. The latter is the kind we use here, which means that  the network 
predicts a numerical value for each of the five different SIE
parameters ($x$, $y$, $e_x$, $e_y$, and $\theta_{\rm E}$).

\begin{figure*}[ht!]
\centering
\includegraphics[trim=0 350 0 350, clip, width=\textwidth]{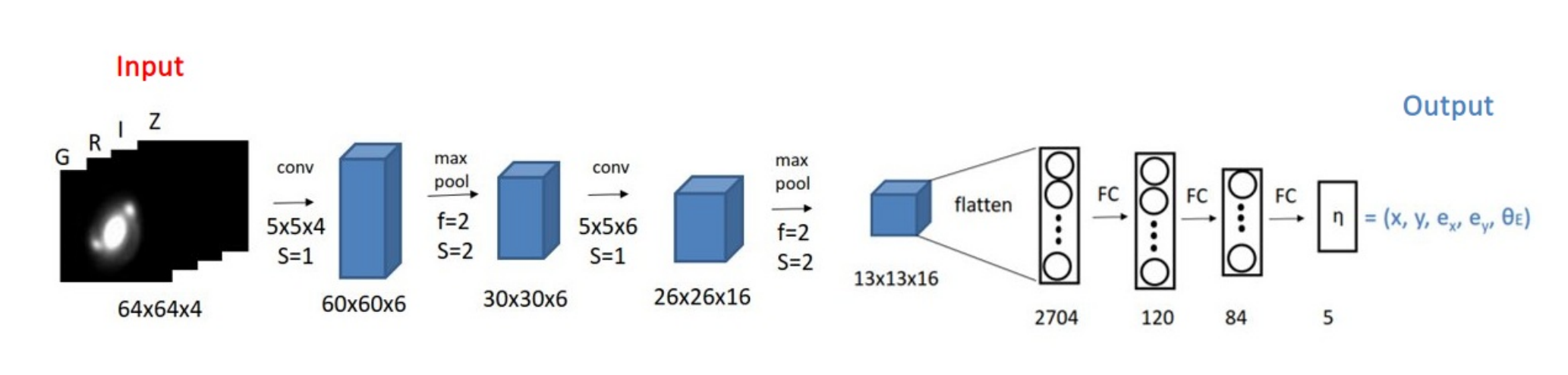}
\caption{Overview of our main CNN architecture. The input has four different filter images for each lens system and each image a size of $64 \times 64$ pixels. The network contains two convolutional layers (conv) each followed by a max-pooling layer (max pool) with kernel size $f$ and stride $S$ values indicated in the figure. This is then followed, after flattening the data cube, by three fully connected (FC) layers to finally obtain the five output values of the SIE $\eta$, containing the lens center $x$ and $y$, the complex ellipticity $e_x$ and $e_y$, and the Einstein radius $\theta_\text{E}$.}
\label{fig:CNNOverview}
\end{figure*}

Depending on the problem the network needs to solve, there are several
different types of networks. Since we
are using images as data input, typically convolutional layers
followed by fully connected (FC) layers are used
\citep[e.g.,][]{hezaveh17, levasseur17, pearson19}. The detailed
architecture depends on attributes such as the specific task,
the size of the images, or the size of the data set. We have tested
different architectures and found an overall good network performance
with two convolutional layers followed by three FC layers, but no significant improvement for the other network architectures. A sketch of
this is shown in Figure~\ref{fig:CNNOverview}. The
input has four different filter images for each lens system and each image a
size of $64\times 64~\text{pixels}$. The convolutional layers have a 
stride of 1 and a kernel size of $5\times5 \times C$, with $C=4$ for the first layer and $C=6$ for the second layer. Each convolutional layer is
followed by a max-pooling layer of size $f \times f=2\times 2$ and stride 2. We use as activation function the rectified linear activation (ReLu) function. After the two
convolutional layers, we obtain a data cube of size
$13\times13\times16$, which is then passed through 
the FC layers after flattening to finally obtain the five output
values. This network is coded with python 3.8.0 and uses pytorch modules (torch 1.5.1).

Independent of the exact network architecture, the network can contain hundreds of thousands of neurons or more. While initially the values of weight parameters and bias of each neuron are random, they are   updated during the training. To see the network performance after the training,  the data set is split into three samples: the training, the validation, and the test sets. We further divide those sets into random batches of size $N$. In each iteration the network predicts the output values for one batch (forward propagation), and after running over all batches from the training and validation sets, one epoch is finished. The error, which is called loss, is obtained for each batch with the loss function;  we use the mean square error (MSE) defined as
\be
L = \frac{1}{p \times N} \sum_{k=1}^N \sum_{l=1}^p (\eta^\text{pred}_{k,l}-\eta^\text{tr}_{k,l})^2 \times w_l \, ,
\label{eq:loss}
\ee
where $\eta_{k,l}^\text{tr}$ and $\eta_{k,l}^\text{pred}$  respectively denotes the $l{th}$  true and predicted parameter, in our case from $ \{ x,y, e_x, e_y, \theta_\text{E}\}$, of lens system $k$, and $p$ denotes the number of output parameters. We incorporated in our loss function $L$ weighting factors $w_l$, which are normalized such that $\sum _{l=1}^p w_l =p$ holds. This gives a weighting factor of 1 for all parameters if they are all weighted equally.

The loss value of that batch is then propagated to the weights and biases (back propagation) for an update based on a stochastic gradient descent algorithm to minimize the loss. This procedure is repeated in each epoch first for all batches of the training set and an average loss is obtained for the whole training set. Afterwards the steps are repeated for all batches of the validation set, while no update of the neurons is done, and an average loss for the validation set is obtained as well. The validation loss shows whether the network improved in that epoch or if a decreasing training loss is related to overfitting the neurons. 
A network is overfitting when it learns the training set, and not the features in the training set. After each epoch we reshuffle our whole training data to obtain a better generalization. This concludes one epoch, which is repeated iteratively to obtain a network with optimal accuracy.

This whole training corresponds to one so-called cross-validation run, where several cross-validation runs are performed by exchanging the validation set with another subset
of the training set.  For
example, if the training set and validation set form five subsets \{A, B,
C, D, and E\}, then we can have five independent runs of training where in
each run the validation set is one of these five subsets and the
training set contains the remaining four subsets.
After
the multiple runs, we can determine the optimal number of epochs for 
training by locating the epoch with the minimum average validation
loss across the multiple runs. This procedure helps to minimize potential bias to certain types of lenses for a potentially unbalanced single split. The neural network trained on all five sets \{A, B, C, D, E\} up to that epoch is the final network, which is then applied to the test set that
contains data the network has never seen before. In our case we used
$\sim$56\% of the data set as the training set, $\sim$14\% as  the validation set, and $\sim$30\%
as the test set\footnote{The percentiles vary slightly due to rounding effects depending on the absolute size of the simulated mocks of that sample and the assumed batch size.} in order to  have a five-fold cross-validation for each
network.

To  find the best hyperparameter values for our specific problem, we test  each network on its
 performance with several different variations of the
hyperparameters. Independent of the data set, we train
each cross validation 
run for 300 epochs, and apart from a few checks with different
values, we fix the weight decay to 0.0005 and the momentum to 0.9. For
the learning rate, batch size, and the initializations of the
neurons, we perform a grid search, varying the learning rate $r_\text{learn} \in [0.1, 0.05, 0.01, 0.008, 0.005, 0.001, 0.0005, 0.0001]$ and
 batch size as
32 or 64 images per batch, and exploring three different network
initializations. For the weighting factors of the contribution to the
loss we test  two options, either all parameters contribute
equally (i.e., $w_l =1~ \forall~ l $ in eq. \ref{eq:loss}) or the
contribution of the Einstein radius is a factor of 5 higher
($w_{\theta_\text{E}} = 5$). This  already gives 96 different
combinations of hyperparameters  which we test with cross-validation
and early stopping.

For a subset of the 
 hyperparameter combinations, we test further possibilities. In
 particular, we explore the effect of drop-out (i.e., the omission of random neurons in every iteration) with a drop-out rate $p \in
[0.1,0.3,0.5,0.7,0.9],$ but find no improvement. We further test
different network architectures by adding an additional convolutional
layer or fully connected layer, or by varying the number of neurons in
the different layers. We also test a different weighting of the lens center parameters to the loss that is motivated by the results of our networks in Section 4.

In addition, we test the effect of five different scaling options of the
input images for our data set,
but assume here the learning rate
$r_\text{learn} = 0.001$ for simplicity. First, we boost the $r$ band by a
factor of 10. Since the network is still able to recover the parameter
values, we see that the network performance is not heavily affected by
the absolute value of the images. Second, if we normalize each filter of one lens
system independently of the other filters, the network fails to
recover the correct parameter values. This shows us that
the network is indeed able to extract the color information
as the relative difference is much smaller, and thus needs the
different filters. In the third and fourth option we normalize  each filter with its maximum value or with the mean
peak value of the different filters. Last, we
also rescale the images by shifting them by the mean value and
dividing by the standard deviation\footnote{The four individual images
  are rescaled as
\be
I_\text{scaled} = \frac{(I - M)}{ \sigma}
\ee
with mean
\be
M = \sum_{k=0}^f\sum_{l,m=0,0}^{p1,p2} I_{k,l,m}/(f\times p1 \times p2) \, ,
\ee
the number of filters $f$, and
\be
\sigma =  \sqrt{\sum_{k=0}^f \sum_{l,m=0,0}^{p1,p2} \frac{(I_{k,l,m} -M)^2 }{(p1 \times  p2-1)}} \, ,
\ee
 and $p1$ and $p2$ as image dimensions in pixels for the $x$-  and $y$-axis, respectively. In our case we have $f=4$ and
  $p1=p2=64$.}. Since we obtained no notable
improvement with any one of
these scalings, we use the images without rescaling to obtain our
final networks.

\FloatBarrier
\section{Results}
\label{sec:results}

To train our modeling network we mock up lensing systems based on
real observed galaxies with our simulation pipeline (see
Sect. \ref{sec:simulation}). Each lensing system is simulated in the
four different filter $griz$ of HSC to give the network color
information to distinguish better between lens galaxy and lensed
arcs. The network architecture assumes, as described in
Sect. \ref{sec:network} in detail, images   64 $\times$ 64
pixels in size, which corresponds to a size of around $10\arcsec\times
10\arcsec$.

During our network testing, we found that the distribution of
Einstein radii in the training set is very important, especially as this is a key
parameter of the model. Therefore, we trained a network under the
assumption of different underlying data sets, for example a lower limit of
the Einstein radius for the simulations or a different distribution of
Einstein radii. We further tested the network performance by limiting
to a specific configuration (i.e., only doubles or quads). We give an
overview of the different data set assumptions in Table \ref
{tab:overview}, as well as the best hyperparameter values that depend on
the assumed data set.

\begin{table*}
\begin{center}
\caption{Overview of trained networks.}
\label{tab:overview}
\begin{tabular}{c|c|c|c||c|c|c|c|c||c|c}
  \multicolumn{11}{c}{Natural distribution of Einstein radii of lenses} \\
  \hline
  double     & quad & $\theta_\text{E,min}$ [\arcsec]  & $w_{\theta_\text{E}}$  & loss  & epoch  & $r_\text{learn}$  & $N$ & seed & Section  & Figures \\
  \hline
  \checkmark & \checkmark &  0.5 & 1    & 0.0201 & 115    & 0.005  & 64 & 3  & \ref{sec:results:normal0p5}        &      \\ 
  \checkmark & \checkmark &  0.5 & 5    & 0.0496 & 123    & 0.001  & 64 & 3  & \ref{sec:results:normal0p5}        & \ref{fig:comparison_normal0p5}, \ref{fig:cornerplot}, \ref{fig:brightness}, \ref{fig:comparison_im_pos_timedelay}, \ref{fig:comparison_rE_timedelay} \\ 
  \checkmark & \checkmark &  2.0 & 1    & 0.0120 & 85     & 0.01   & 32 & 3  & \ref{sec:results:normal2p0}        &      \\ 
  \checkmark & \checkmark &  2.0 & 5    & 0.0209 & 85     & 0.008  & 32 & 2  & \ref{sec:results:normal2p0}        & \ref{fig:cornerplot}, \ref{fig:brightness}, \ref{fig:comparison_normal2p0}, \ref{fig:comparison_im_pos_timedelay}, \ref{fig:comparison_rE_timedelay} \\ 
 \hline
  \checkmark &            &  0.5 & 1    & 0.0193 & 242    & 0.008  & 64 & 1  & \ref{sec:tests:doubleorquadsonly}  &       \\ 
  \checkmark &            &  0.5 & 5    & 0.0474 & 117    & 0.001  & 64 & 3  & \ref{sec:tests:doubleorquadsonly}  &       \\ 
  \checkmark &            &  2.0 & 1    & 0.0118 & 163    & 0.05   & 64 & 3  & \ref{sec:tests:doubleorquadsonly}  &       \\ 
  \checkmark &            &  2.0 & 1    & 0.0118 &  96    & 0.01   & 32 & 2  & \ref{sec:tests:doubleorquadsonly}  &       \\ 
  \checkmark &            &  2.0 & 5    & 0.0217 &  62    & 0.008  & 32 & 3  & \ref{sec:tests:doubleorquadsonly}  &       \\ 
 \hline
             & \checkmark &  0.5 & 1    & 0.0193 & 151    & 0.008  & 32 & 2  & \ref{sec:tests:doubleorquadsonly}  &       \\ 
             & \checkmark &  0.5 & 5    & 0.0441 & 69     & 0.001  & 32 & 2  & \ref{sec:tests:doubleorquadsonly}  &       \\ 
             & \checkmark &  2.0 & 1    & 0.0129 & 267    & 0.01   & 64 & 2  & \ref{sec:tests:doubleorquadsonly}  &       \\ 
             & \checkmark &  2.0 & 5    & 0.0268 & 285    & 0.005  & 32 & 1  & \ref{sec:tests:doubleorquadsonly}  &       \\ 
 \hline
  \multicolumn{11}{c}{ } \\
 \multicolumn{11}{c}{Uniform distribution of Einstein radii of lenses} \\
 \hline
 double     & quad & $\theta_\text{E,min}$ [\arcsec]  & $w_{\theta_\text{E}}$  & loss  & epoch  & $r_\text{learn}$  & $N$ & seed & Section & Figures\\
 \hline
  \checkmark & \checkmark &  0.5 & 1    & 0.0223 & 147    & 0.001  & 32 & 1  & \ref{sec:results:equally}         &   \\ 
  \checkmark & \checkmark &  0.5 & 5    & 0.0528 & 112    & 0.0005 & 64 & 2  & \ref{sec:results:equally}         & \ref{fig:cornerplot}, \ref{fig:brightness}, \ref{fig:comparison_equally}, \ref{fig:loss_equally}, \ref{fig:comparison_im_pos_timedelay}, \ref{fig:comparison_rE_timedelay} \\ 
 \hline
             & \checkmark &  0.5 & 1    & 0.0288 & 73     & 0.008  & 64 & 2  & \ref{sec:tests:doubleorquadsonly} &   \\ 
             & \checkmark &  0.5 & 5    & 0.0688 & 56     & 0.001  & 32 & 2  & \ref{sec:tests:doubleorquadsonly} & \ref{fig:comparison_quadsonly} \\ 
\end{tabular}
\end{center}
Note. The first and second columns indicate if quads and/or
doubles are included in the data set. The parameter $\theta_\text{E,min}$ represents
the lower limit on the Einstein radius in the simulation, and
$w_{\theta_\text{E}}$ is the weighting factor of the Einstein radius
in the loss function. The other parameters (lens center,
ellipticity) are always weighted by a factor of 1 and the sum of all five weighting factors is normalized to the number of parameters. The fifth and sixth
columns give the value of the loss of the test set and the epoch with the best
validation loss. This is followed by the specific hyperparameters:
learning rate $r_\text{learn}$, batch size $N$, and seed for the random number generator. The last two columns list the sections and the figures that present the results of the corresponding network.
\end{table*}

We present in the following subsections our CNN modeling results for various data sets.

\subsection{Naturally distributed Einstein radii with lower limit $0.5\arcsec$}
\label{sec:results:normal0p5}

\begin{figure}[ht!]
\includegraphics[trim=20 80 120 0, clip, width=0.45\textwidth]{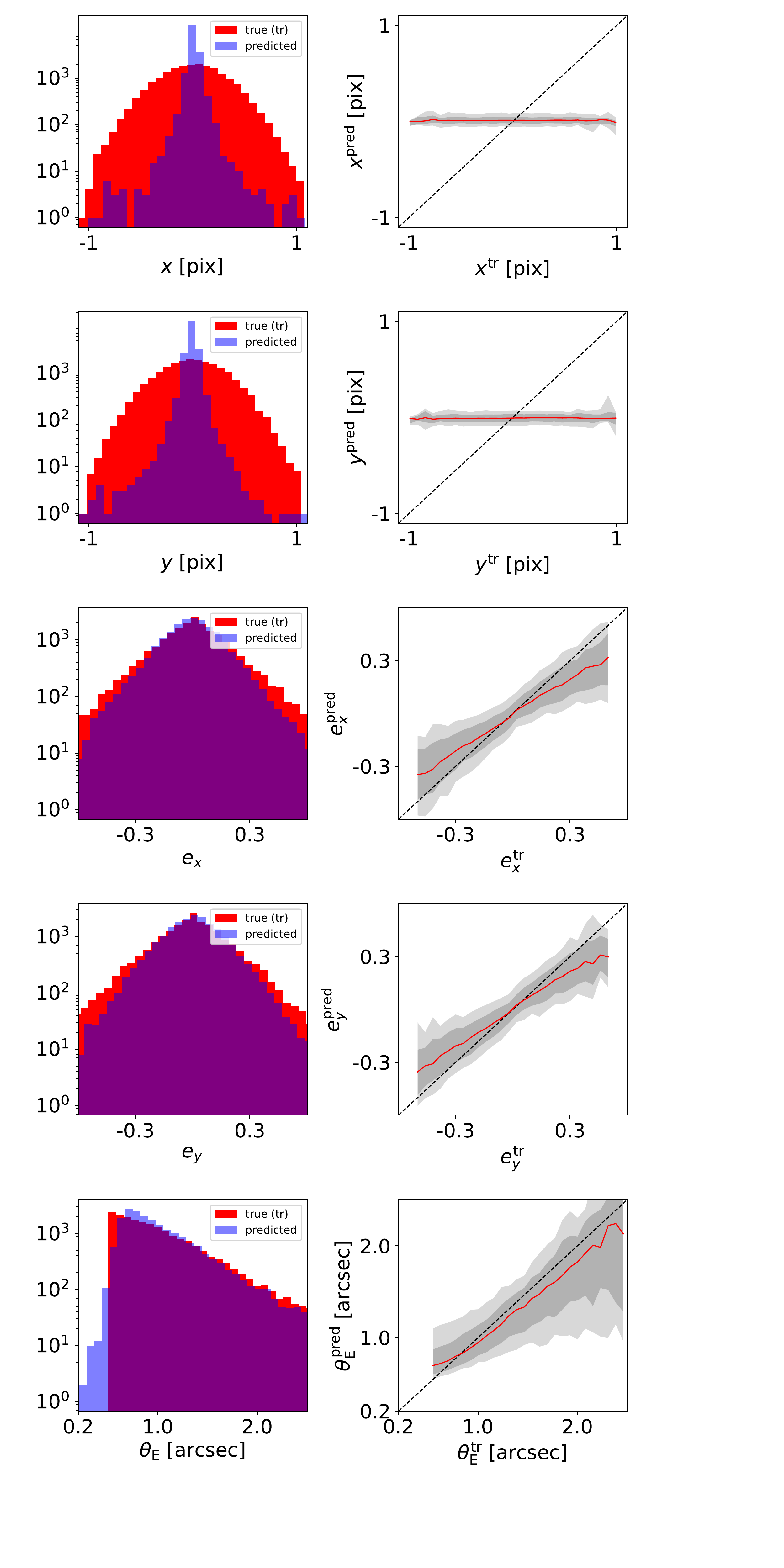}
\caption{Network performance on the Einstein radius under the
  assumption of a lowest Einstein radius $\theta_\text{E,min}$ of
  0.5\arcsec and a weighting factor of $w_{\theta_\text{E}}=5$. The left panel shows histograms of the ground truth (tr) in red and of the predicted values in blue. The right panel
    shows a direct comparison of the predicted against the true value, where the red
  line indicates the median of the distribution and the gray bands give
  the 1$\sigma$ (16$\text{th}$ to 84$\text{th}$ percentile) and 2$\sigma$ ( 2.5$\text{th}$ to 97.5$\text{th}$ percentile) ranges. From top
  to bottom are the five different model parameters, lens center $x$
  and $y$, complex ellipticity $e_x$ and $e_y$, and Einstein radius
  $\theta_\text{E}$. For all plots 30 bins over the plotting range are used.}
\label{fig:comparison_normal0p5}
\end{figure}

For this network we use 65,472 mock lens images simulated following the procedure
described in Sect. \ref{sec:simulation}. Here we assume a lower limit
of the Einstein radii of $0.5\arcsec$ as otherwise the lensed source
is totally blended with the lens and is not resolvable given the average seeing and image quality. The resulting redshift distributions are shown as the blue histograms for the lens in Figure~\ref{fig:hist_lens} (top panel) and for the source in Figure~\ref{fig:hist_zs}. The lens redshift peaks at $z_\text{d} \sim 0.5$. Concerning the possible strong lensing
configurations, the data set is dominated by doubles as expected. In addition, systems with smaller
Einstein radii are more numerous than those with larger Einstein
radii, as expected given the lens mass distribution, although the velocity dispersion (see  Figure~\ref{fig:hist_lens}, bottom panel) peaks at around $v_\text{disp} \sim 280 \,\text{km}\, \text{s}^{-1}$, and thus tends to include more massive galaxies than the input catalog (gray histogram). The distribution of the different parameters are shown in Figure \ref{fig:comparison_normal0p5},  left panel; the red histogram depicts the true distribution and the blue one the predicted distribution. In the right panel we show the correlation between the true value and the predicted value for the five different parameters.

If we look at the performance on the lens center, which is measured in units of pixels
with respect to the image cutout center, it seems as if the network fails
totally in the first instance. We recall here   how we obtain the lens mass center. In the simulation, we
assume the lens \textit{light} center to be the image center, and add
a Gaussian variation (with standard deviation of $0.05\arcsec$) to shift to the lens \textit{mass} center. Thus, the ground truth (red histogram in
Figure \ref{fig:comparison_normal0p5}) follows a Gaussian distribution,
while the predicted lens center distribution (blue) is peakier. 
This suggests that the network does not obtain enough information
  from the slight shift or distortion in the lensed arcs to 
  correctly predict the lens mass center.
We test upweighting the contribution of the lens
center to the loss with a higher fraction, which results in a better
performance on these two parameters, but then the performance on the
other parameter deteriorates. We thus refrain from upweighting the lens center.
  Further difficulties on the centroid parameters are caused by
   all systems having the exact same lens light
  center (which is at the center of the image). If we   assume that the lens mass perfectly follows the
  light distribution and the lens light center is always the same, the
  lens (mass) center ground truth will become a delta distribution,
  and the network will perform much better.
  Accordingly, in many
  automated lens modeling architectures \citep[e.g.,][]{pearson19} the lens center is not even predicted. Since the
  difference of the center  for nearly all lens systems is smaller than
  $\pm1$ pixel, it does not affect the model noticeably. We nonetheless keep five parameters for generality, and 
   suggest  investigating in future work more in this
  direction by relaxing the strict assumption of coincidence centers
  of image cutout and of lens light.

Looking at the performance on the ellipticity, it turns out
that most of the lens systems are approximately round (i.e., $e_x \sim e_y \sim 0$)
and that the network can recover them very well. If the lens is
more elliptical, the network performance starts to drop. This might be
an effect of the lower number of such lens systems in the sample especially
since the position angle becomes relevant, and thus the number of
systems in a particular direction is again lower.
We note that $e_x= \pm 0.3$ and $e_y=0$ corresponds to an axis ratio $q=0.73$ (i.e., quite elliptical). If the absolute value of $e_x$ or $e_y$ were
  higher, the axis ratio would be even lower, which seldom occurs   in nature.

We see that the network recovers the Einstein radius better for lens
systems with lower image separation than with high image separation ($\theta_\text{E}\gtrsim 2\arcsec$), which is in the first instance
counter-intuitive. If the lensed images are further separated, they are better resolved and less strongly blended with the lens, and we would expect
better recovery of Einstein radii from the network. The worse network
performance at larger Einstein radii can therefore only be
explained by the relatively low numbers of these systems in the training data. We have more 
than two orders of magnitude more lens systems with $\theta_\text{E} \sim 0.5
\arcsec $ than with $\theta_\text{E} \sim 2.0 \arcsec $. Therefore, the
network is trained to predict  a small Einstein radius more often,  and a larger Einstein radius less often. Since
the lens systems with larger image separation are very interesting for
a wide range of scientific applications, it is desirable to improve the network
performance  specifically on those lens systems. Therefore, we test a
network with the same data set where the Einstein
radius difference contributes a factor of 5 more to the loss
than the other parameters. In the case of this weighted network, the
prediction performance is very similar for the lens center and
ellipticity, but slightly better for the Einstein radius. If we
increase the contribution of the Einstein radius further, we notably worsen
 the performance on the other parameters.

\begin{figure*}[ht!]
\includegraphics[trim=0 0 0 0, clip, width=1.0\textwidth]{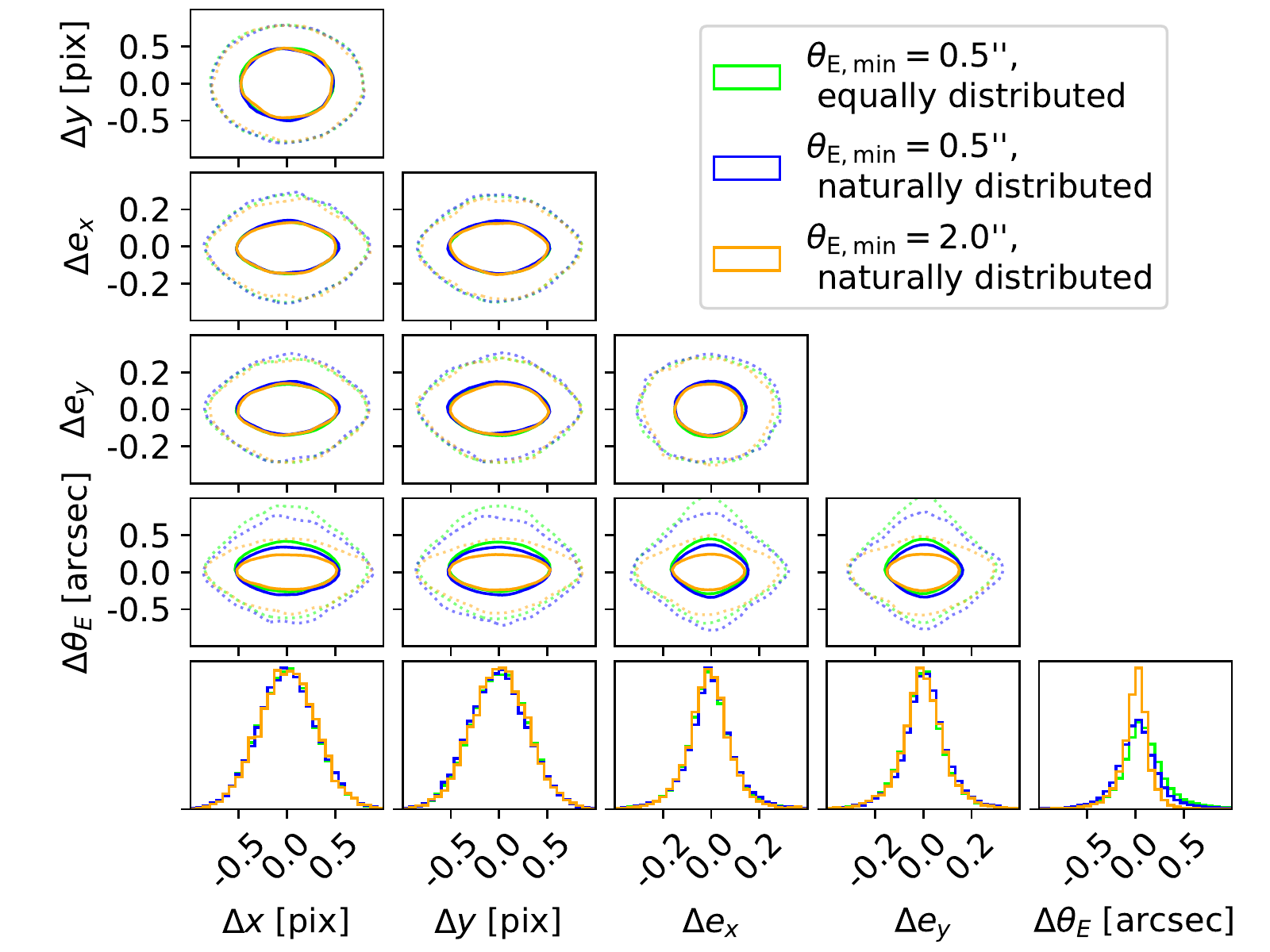}
\caption{Comparison of the performance of the three networks described in Sect. \ref{sec:results}. All samples include doubles and quads and a weighting factor of $w_{\theta_\text{E}}=5$, but different Einstein radius distributions or lower limits on the Einstein radius (see legend). In the bottom row are  shown the normalized histograms of the difference between predicted values and ground truth for the five parameters and above the 2D correlations distribution: 1$\sigma $ contour (solid line) and   2$\sigma$ contour (dotted line).
}
\label{fig:cornerplot}
\end{figure*}

As a further comparison of the ground truth with the predicted values
of the test set, we show in Figure \ref{fig:cornerplot} the difference
as normalized histograms (bottom row) and the 2D probability
distributions (blue),  where we find no strong correlation among the
five parameters. The obtained median values with 1$\sigma$
uncertainties for the different parameters are, respectively,
$(0.00^{+ 0.31 }_{ -0.30 })\arcsec$ for $\Delta x$,
$(-0.01^{+ 0.29 }_{  -0.31 }) \arcsec$ for $\Delta y$,
$ 0.00 ^{+ 0.08 }_{ -0.09 }$ for $\Delta e_\text{x}$,
$ 0.01 ^{+ 0.09 }_{ -0.08 }$ for $\Delta e_\text{y}$, and
$ (0.02 ^{+ 0.21 }_{ -0.18 }) \arcsec$ for $\Delta \theta_\text{E}$,
where $\Delta$ denotes the difference between the
predicted and ground truth values. As an example, a shift of $e_x=0.3$
to $e_x=0.15$ with fixed $e_y=0$ results in a shift from $q=0.73$
to $q=0.86$. 

\begin{figure*}[ht!]
\includegraphics[trim=0 0 0 0, clip, width=0.5\textwidth]{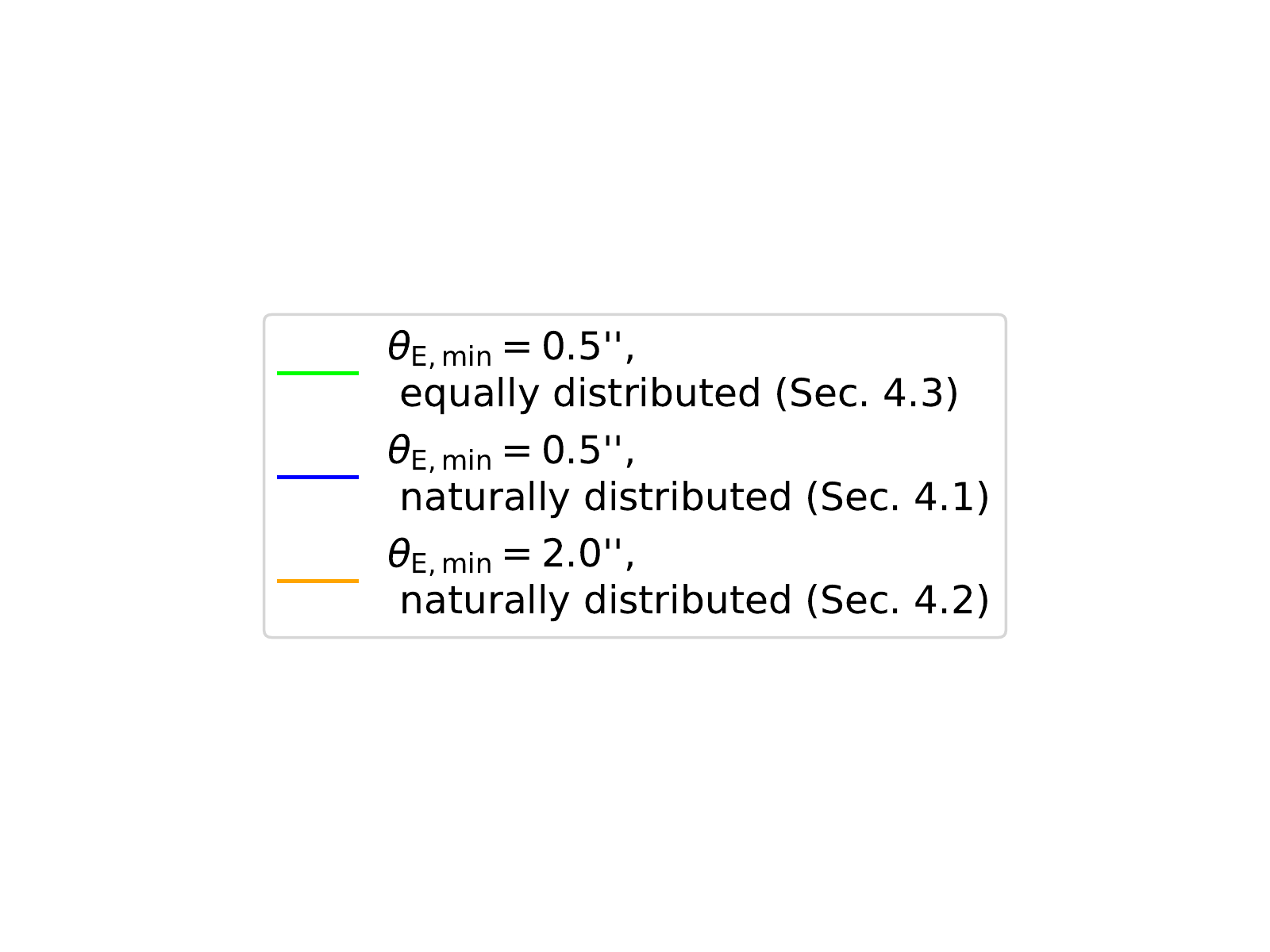}
\includegraphics[trim=0 0 0 0, clip, width=0.5\textwidth]{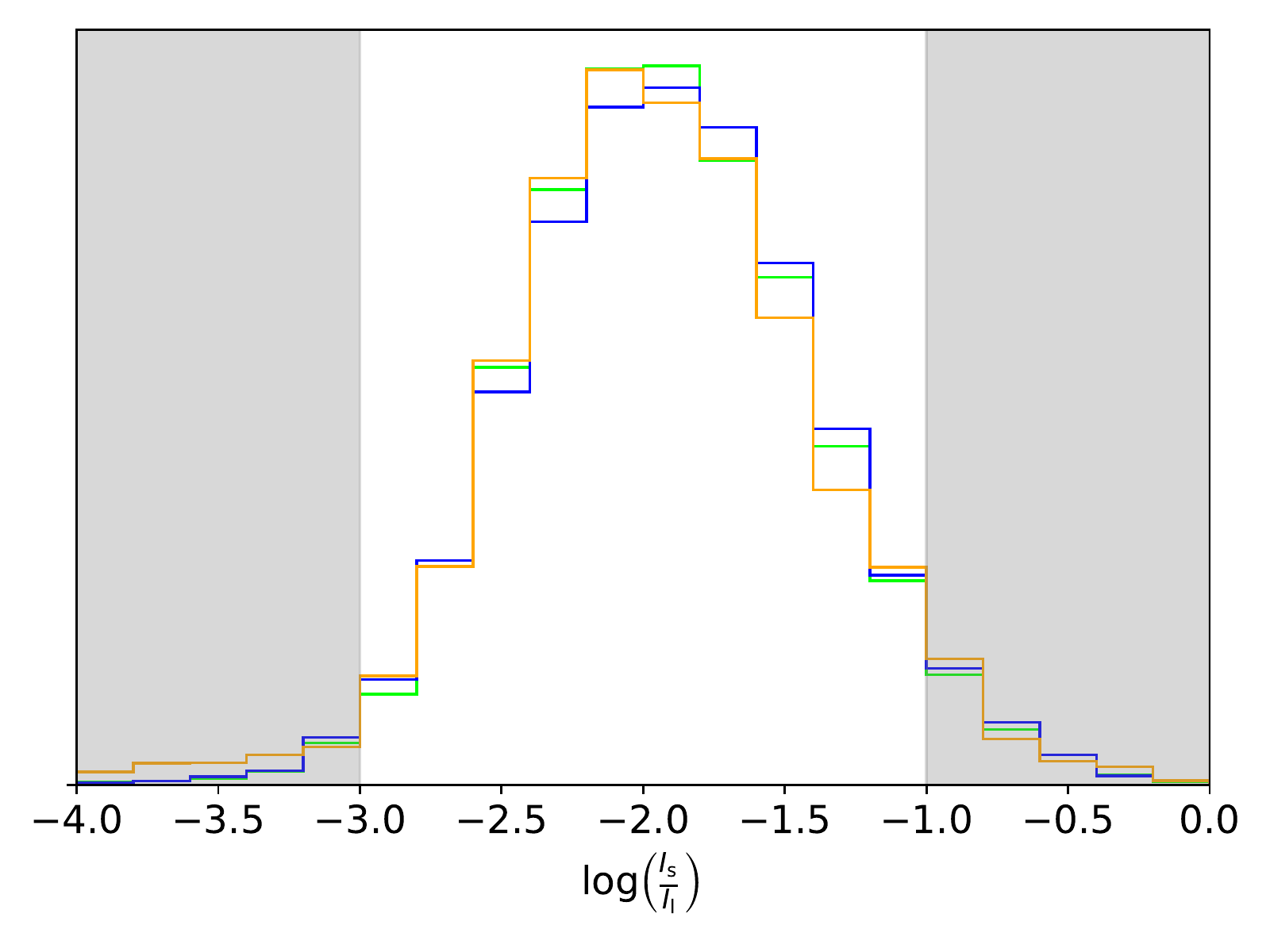}

\includegraphics[trim=0 0 0 0, clip, width=0.5\textwidth]{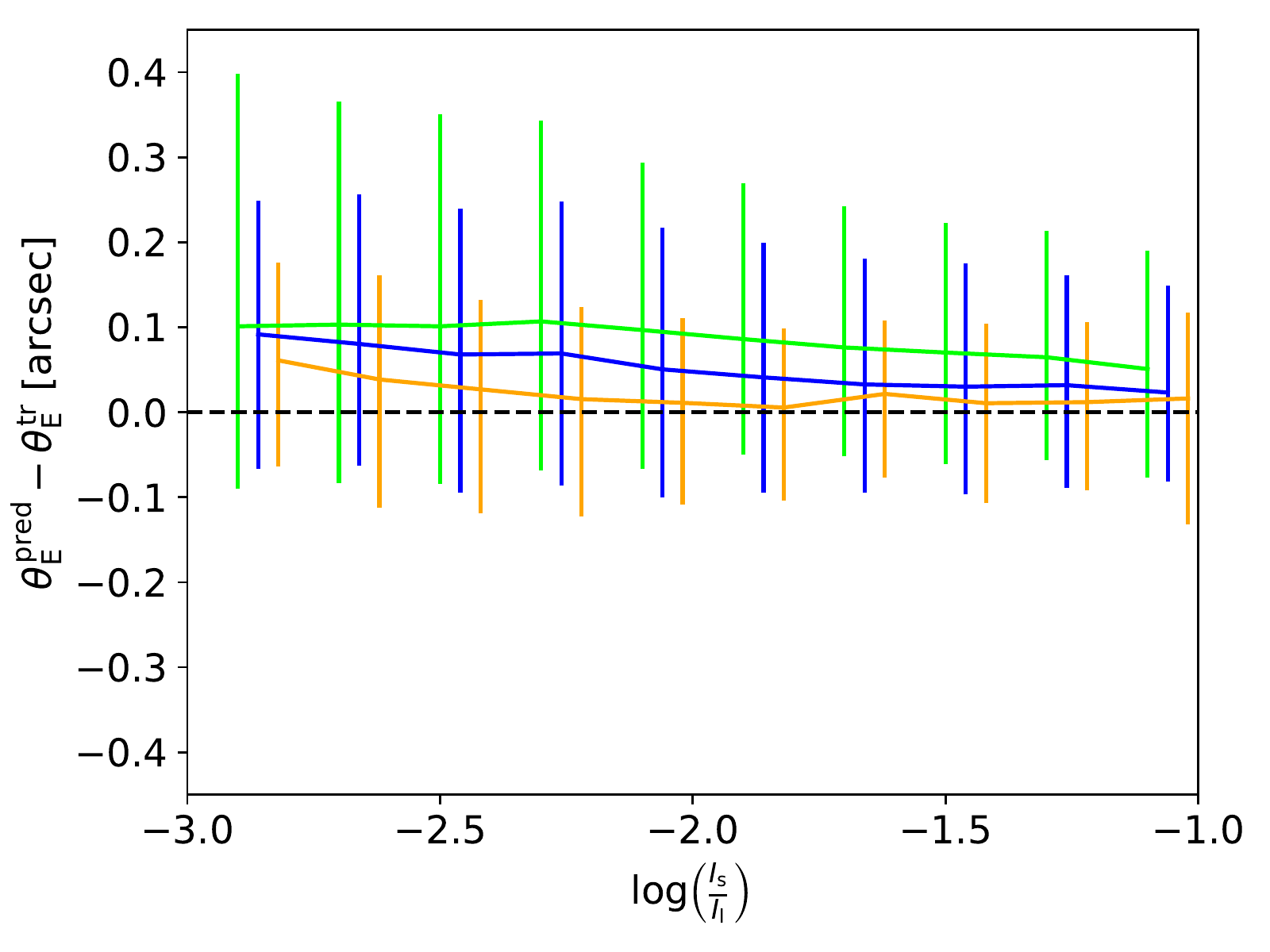}
\includegraphics[trim=0 0 0 0, clip, width=0.5\textwidth]{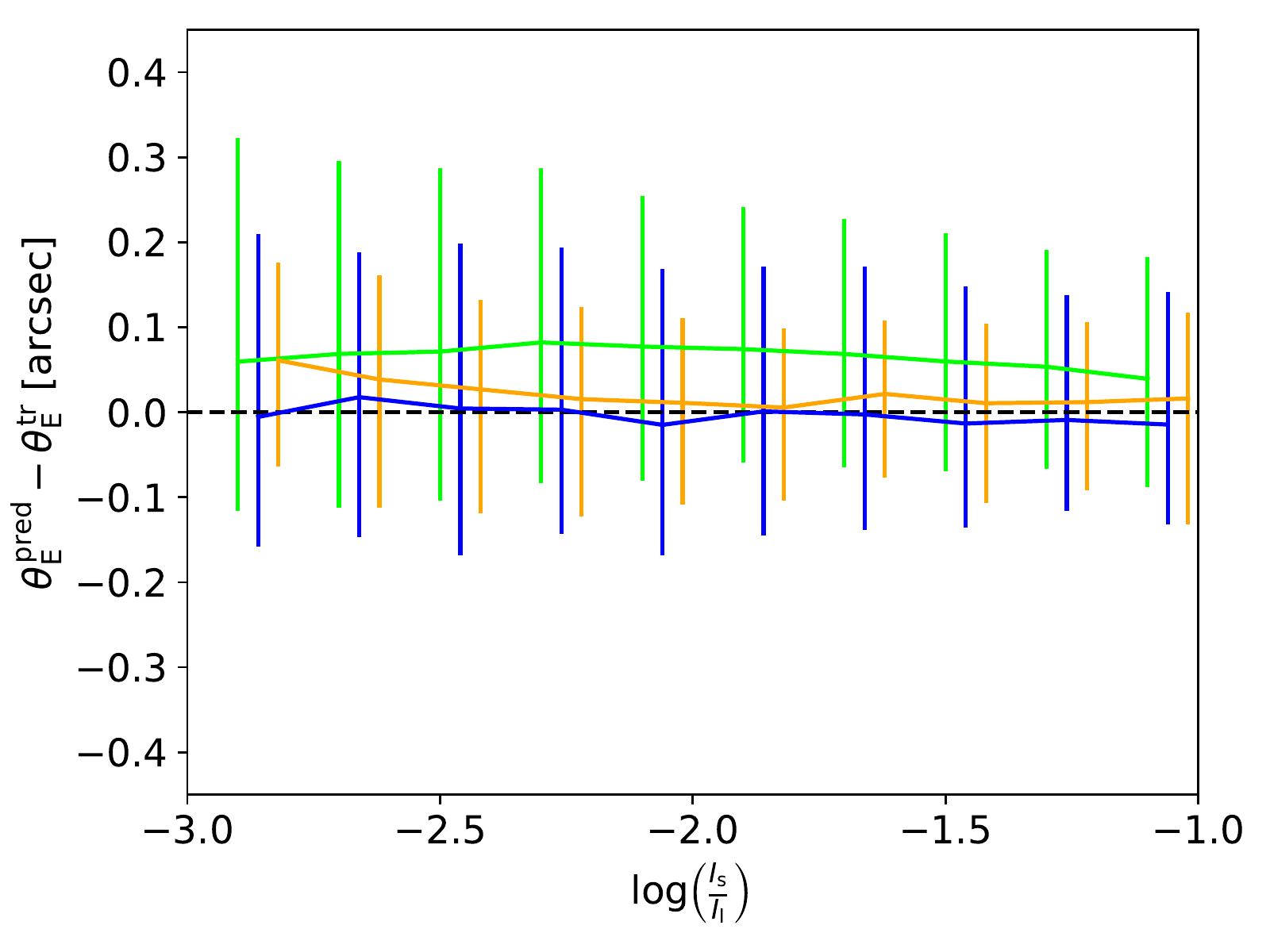}
\caption{Comparison of the performance of the three networks described in Sect. \ref{sec:results}. All samples include doubles and quads and a weighting factor of $w_{\theta_\text{E}}=5$, but different Einstein radius distribution or lower limits on the Einstein radius as indicated in the legend (upper left). The upper right panel shows the histogram of the brightness ratio of lensed source and lens. The bottom panel shows for the full sample (left) and limited to $\theta_\text{E} > 0.8\arcsec$ (right) the difference in Einstein radius as a function of the brightness ratio with the 1$\sigma$ values. Shown are the Einstein radius difference in the range $-3 \leq \text{log} \left(\frac{I_s}{I_l} \right) \leq -1$ (white area in the histogram) where there are enough data points, and  the blue and orange bars have been shifted slightly to the right  for better visualization.}
\label{fig:brightness}
\end{figure*}

Finally, we show in Figure~\ref{fig:brightness} the difference in
Einstein radii as a function of the logarithm of the ratio between
lensed source intensity $I_\text{s}$ and lens intensity $I_\text{l}$
determined in the $i$ band, which we hereafter refer to as the
brightness ratio. In the top right panel, we show the distribution of the
brightness ratio. The lens intensity is defined as the sum of all the
pixel values in the 64 pixels $\times$ 64 pixels cutout of the lens
such that it is slightly overestimated due to light contamination from
surrounding objects. The distribution peaks around $-2$ in
  logarithm to basis 10, which means that the lensed source flux is a
factor 100 below that of the lens. The bottom left plot shows the
  median with 1$\sigma$ values of the Einstein radius differences for
  each brightness ratio bin. Focussing on the blue curve for this section, we find a bias in the Einstein
  radius which is driven by the small lensing systems with
  $\theta_\text{E} \lesssim 0.8\arcsec$ (compare Figure
  \ref{fig:comparison_normal0p5}). Excluding these small lensing systems,
  we show the corresponding plot in the lower right panel. With
  this limitation, we no longer find a bias,  and obtain a median with 
  1$\sigma$ values of $ 0.00_{-0.14}^{+0.17}$$\arcsec$ for the Einstein
  radius difference. We find a slight improvement of the performance
  with increasing brightness ratio for both the full sample (bottom left panel) and the sample with $\theta_\text{E} > 0.8\arcsec$ (bottom right panel).

To further improve the network performance for wide-separation lenses,
we train separate networks for lens systems with Einstein radius $\theta_\text{E}>2.0 \arcsec $ in
Sect.~\ref{sec:results:normal2p0}, and for lens systems where we artificially boost the
number of lenses at the high end of $\theta_\text{E}$ in
Sect.~\ref{sec:results:equally}.


\subsection{Naturally distributed Einstein radii with lower limit $2.0\arcsec$}
\label{sec:results:normal2p0}

\begin{figure}[ht!]
\includegraphics[trim=20 60 100 1100, clip, width=0.5\textwidth]{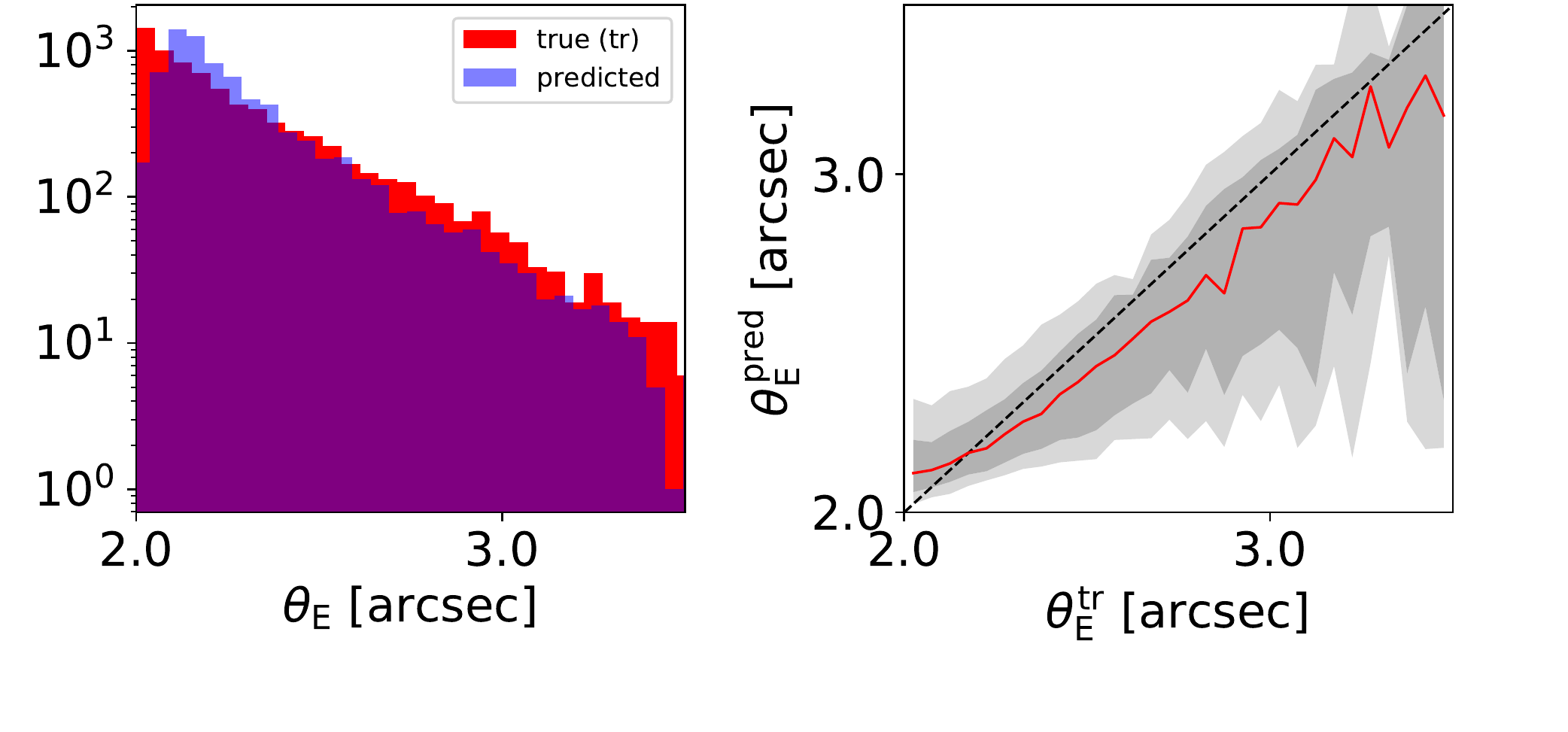}
\caption{Network performance on the Einstein radius under the
  assumption of a smalles Einstein radius $\theta_\text{E,min}$ of $2.0
  \arcsec$. The left panel shows the histograms of the ground truth (tr)
  in red and of the predicted values in blue. The right panel
  is a 1:1 plot of predicted against true Einstein radius. The red
  line shows the median of the distribution and the gray bands give
  the 1$\sigma$ and 2$\sigma$ ranges. For both plots 30 bins over the plotting range are used.}
\label{fig:comparison_normal2p0}
\end{figure}

Since the network presented in Sect.~\ref{sec:results:normal0p5} cannot easily recover a large Einstein radius ($\theta_\text{E}\gtrsim 2\arcsec$), we test the performance of a network specialized for
the high end of the distribution and set the lower limit to
$\theta_\text{E,min}=  2\arcsec$. Because of the higher limit on the Einstein radii, the velocity dispersion (see  orange histogram in Figure~\ref{fig:hist_lens}, bottom) is shifted towards the high end, which corresponds to more massive galaxies. We also find that the lens and source redshifts (orange histograms in Figure~\ref{fig:hist_lens} and Figure~\ref{fig:hist_zs}, respectively) tend to slightly higher values. Since we use the natural
distribution of Einstein radii as in
Sect.~\ref{sec:results:normal0p5}, the image-separation distribution is again bottom-heavy and the number of mock lens systems is smaller (25,623), as shown in Figure~\ref{fig:comparison_normal2p0}. From
the blue (predicted) histogram, we see that the true distribution (red histogram) is
well recovered.

In the right panel of Figure \ref{fig:comparison_normal2p0} we show
the correlation of predicted and true Einstein radii. The red line, which follows quite well the
diagonal dashed line, shows the median. The gray shaded regions
show the $1\sigma$ and $2 \sigma$ regions. We find that the network
performs much better for $\theta_\text{E} \sim 2\arcsec$ than for the
network trained in the full range
(Sect.~\ref{sec:results:normal0p5}). However, this is again due to
the  number of lens systems that decreases towards $\theta_\text{E}\sim
4\arcsec$, and the scatter that increases dramatically for the high end of the
data set.

We further show 1D and 2D probability distributions for this network
in Figure~\ref{fig:cornerplot} (orange) as well as the histogram
  of the brightness ratio, and the difference of the Einstein radii
as function of the brightness ratio in
Figure~\ref{fig:brightness}. While the performance for the lens center
and complex ellipticity is very similar to the network presented in
Sect.~\ref{sec:results:normal0p5}, we achieve an improvement for the
Einstein radius. This is expected as the network is specifically
trained for lens systems with large image separation. As we see from
Figure~\ref{fig:brightness}, the larger systems do not have a higher
brightness ratio on average as one might expect. As we have already seen,
the network performs notably better on the Einstein radii over the
whole brightness ratio range. We no longer overpredict the Einstein
radius for $\log \left( \frac{I_\text{s}}{I_\text{l}} \right) \gtrsim
-2.5$, and  the 1$\sigma$ values are smaller as well.

\subsection{Uniformly distributed Einstein radii with lower limit $0.5\arcsec$}
\label{sec:results:equally}

Because of the extreme decrease in the number of systems towards large image
separation, we test a network trained on a more uniformly distributed
sample. For this, we generate more lens systems with high image
separation by rotating the lens image by $n \pi/2$ with $n \in[0,1,2,3]$. Here we do not reuse the same lens in the same rotation to
avoid producing multiple images of lens systems that are too similar.
We note that the background source and
position are always different such that the lensing effect varies (see
Sect. \ref{sec:simulation} for further details on the simulation
procedure). We limit the sample 
to a maximum of 8,000 lens systems per $0.1\arcsec$ bin resulting in a sample of 140,812 lens systems. This results in a more uniform distribution, though the bins with the largest image separation   still have fewer lens systems since it is very difficult, and  a very seldom occurrence, to obtain a lensing configuration with an image
separation above $\sim$$2.5\arcsec$ due to the mass distribution of galaxy-scale lenses. The biggest image
separation within this sample is $\sim$$4.5\arcsec$, which is about $10\%$ lower than the upper limit of 5$\arcsec$ that we set for our simulations (see Sect.~\ref{sec:simulation:mock}). The redshift distributions, shown as green histograms in Figure~\ref{fig:hist_lens} and Figure~\ref{fig:hist_zs}, are similar to that of the naturally distributed sample (blue), whereas the lens velocity dispersions (Figure~\ref{fig:hist_lens}, bottom panel) tend to be higher (i.e., more massive galaxies), as expected.

\begin{figure}[ht!]
\includegraphics[trim=20 80 120 0, clip, width=0.47\textwidth]{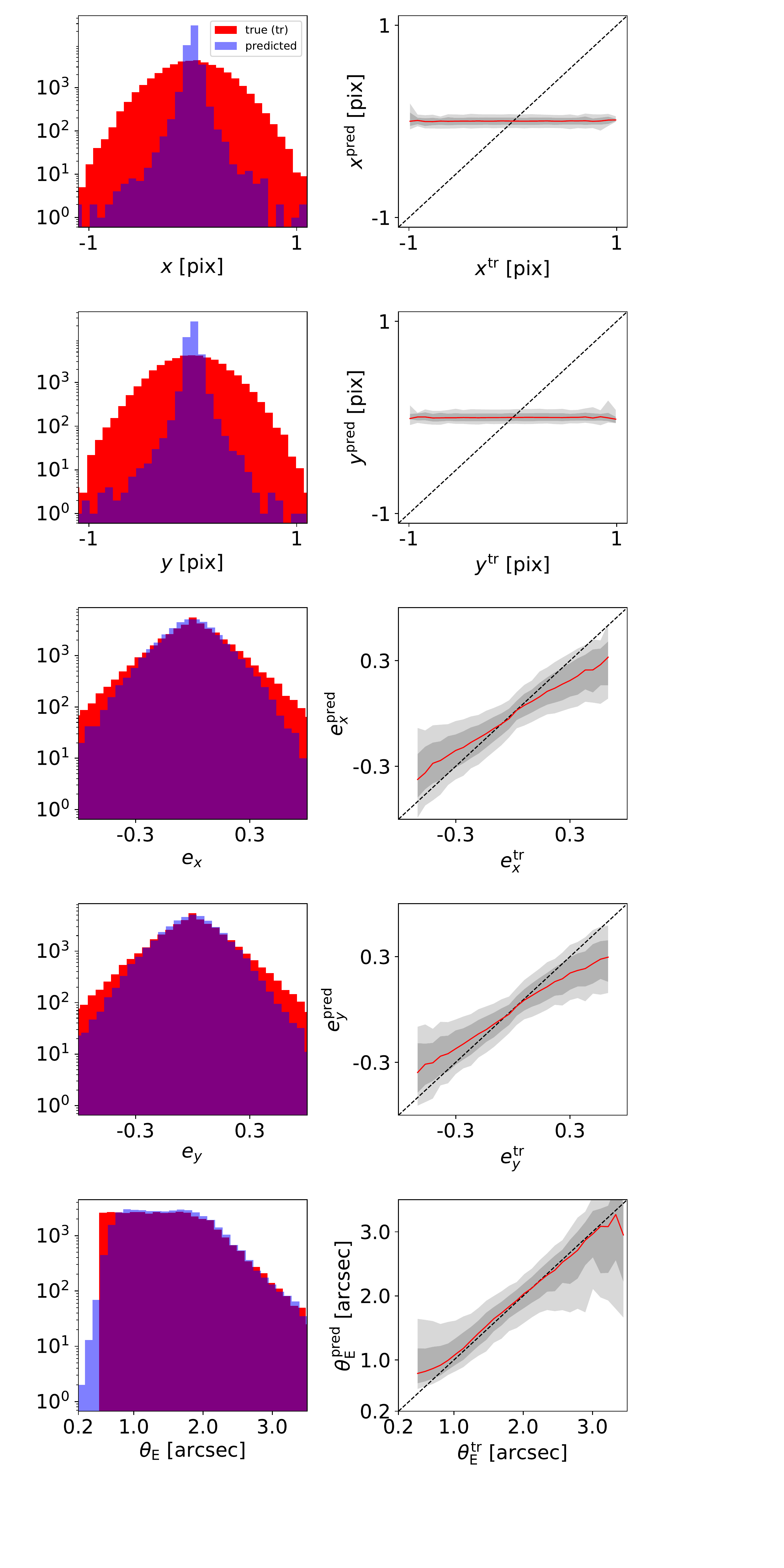}
\caption{Network performance under the assumption of a lowest Einstein
  radius $\theta_\text{E,min}$ of $0.5\arcsec$ but a uniform
  distribution up to $\sim 2\arcsec$. The left panel shows histograms of the ground truth
  (tr) in red and of the predicted values in blue. The right panel
   shows a direct comparison of the predicted against the true value. From top
  to bottom are the five different model parameters, lens center $x$
  and $y$, complex elliptcity $e_x$ and $e_y$, and Einstein radius
  $\theta_\text{E}$. For all plots 30 bins over the plotting range are used.}
\label{fig:comparison_equally}
\end{figure}

Similar to the networks trained with natural Einstein
radius distribution (see Sect. \ref{sec:results:normal0p5} and
Sect. \ref{sec:results:normal2p0}),  in
Figure \ref{fig:comparison_equally} we show histograms (left column) and a 1:1
comparison (right column), but now for all five SIE parameters (from
top to bottom for the lens center $x$ and $y$, the complex ellipticity
$e_x$ and $e_y$, and the Einstein radius $\theta_\text{E}$). For this network we obtain a median value with 1$\sigma$ scatter of
$  (0.00 ^{+ 0.30 }_{ -0.30 }) \arcsec$ for $\Delta x$,
$  (0.00 ^{+ 0.30 }_{ -0.29 }) \arcsec$ for $\Delta y$,
$  -0.01 ^{+ 0.08 }_{ -0.09 }$ for $\Delta e_\text{x}$,
$  0.00 ^{+ 0.08 }_{ -0.09 }$ for $\Delta e_\text{y}$, and
$  (0.07 ^{+ 0.29 }_{ -0.12 })\arcsec$ for the Einstein radius $\Delta \theta_\text{E}$. 
Comparing the performance on the Einstein radius to the network from
Sect. \ref{sec:results:normal0p5} with a natural Einstein radius distribution, we see
a significant improvement for the systems with larger image
separation. Therefore, we can confirm that the underprediction of the
Einstein radius in Sect.~\ref{sec:results:normal0p5} is due to the relatively small number of large-$\theta_\text{E}$ systems in the training
data. On the other hand, based on this plot the new network seems to be slightly
worse for the low-image separation systems. It tends to overpredict the Einstein
radius at $\theta_\text{E}
\lesssim 2.0\arcsec$ such that when we limit to $\theta_\text{E}
> 0.8\arcsec$ as in Sect.~\ref{sec:results:normal0p5}, we only get  a slight improvement in reducing the scatter and obtain $\Delta \theta_\text{E} = (0.07_{-0.08}^{+0.25}) \arcsec$.  Therefore, it turns out that the performance depends sensitively on the training
data distribution. 

We find a very similar performance on the lens center and ellipticity as for the network with the natural distribution (see Sect.~\ref{sec:results:normal0p5}). This is expected since the only difference is the distribution in Einstein radii. This can be further visualized with the 1D and 2D probability contours in Figure~\ref{fig:cornerplot}
(green) that show that overall this network performs  very similarly
  to the network trained on the naturally distributed sample
(blue). For all three networks we find minimal correlation between the
different parameters.

In analogy to the previously presented networks, we show in
Figure~\ref{fig:brightness} the histogram of the brightness ratio and
the Einstein radius differences as function of the brightness ratio
for this network. While the distribution matches that from the sample
with naturally distributed Einstein radius, we overpredict the
Einstein radius more than before. This is related to the
overprediction at smaller Einstein radii (see
Figure~\ref{fig:comparison_equally}), which comes from weighting higher
the fraction of systems with larger image separation. We still
underestimate the Einstein radius at the very high end, as already
noted, but this is negligible for the overall performance compared to
the amount of overestimated systems as we still have a factor of $\sim
$ 100 more of them in our sample. This is the reason why the network
tends to overpredict more strongly than that trained on the naturally
distributed sample (Sect.~\ref{sec:results:normal0p5}, and blue lines in
Figure~\ref{fig:cornerplot} and Figure~\ref{fig:brightness}).

\begin{figure}[ht!]
\includegraphics[trim=0 0 0 0, clip, width=0.5\textwidth]{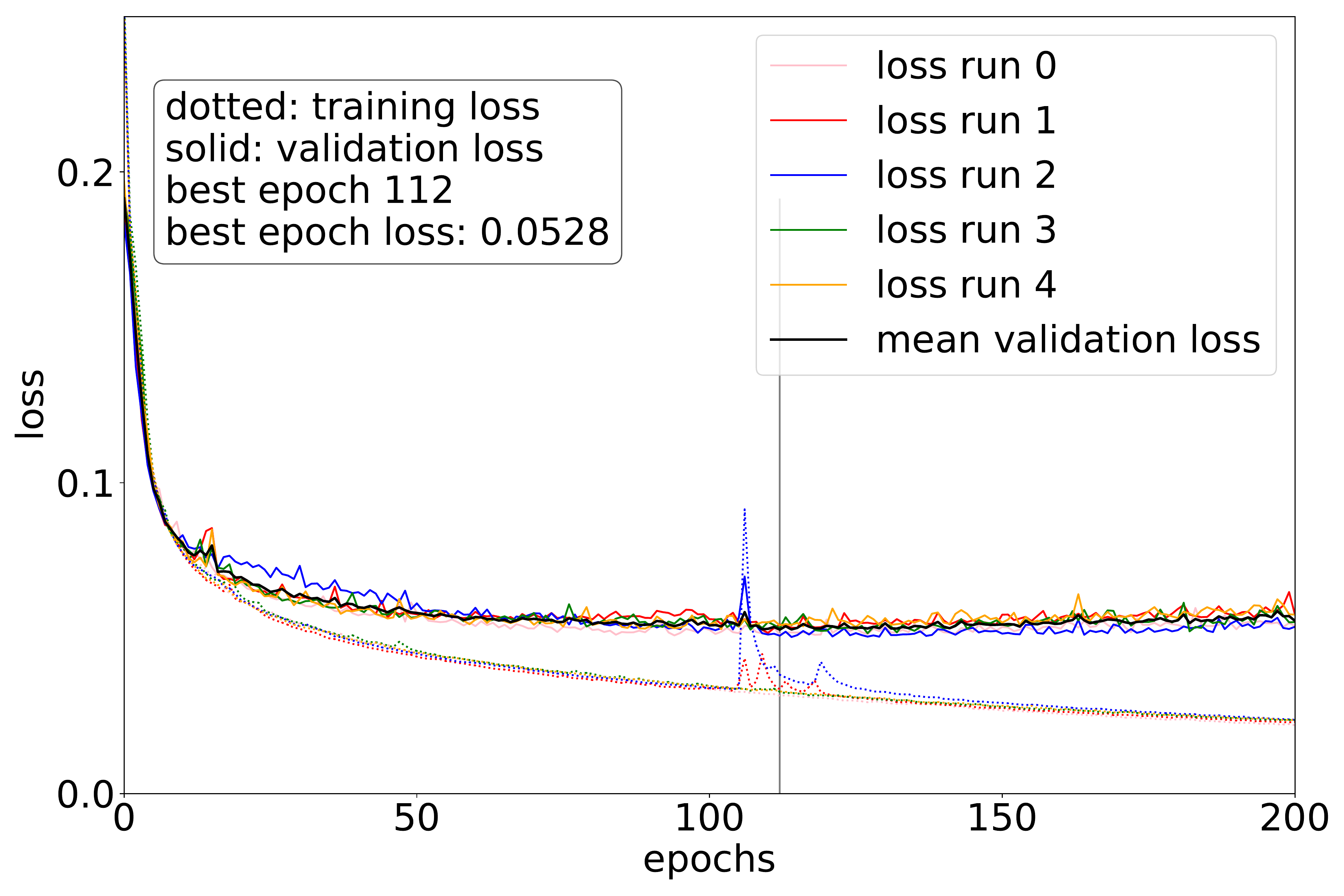}
\caption{Loss curve of our best network under the assumption of equally  distributed Einstein radii. The training loss is shown as dotted lines in   five different colors for the five different cross-validation runs. In the same colors  the validation loss is shown as solid lines together with the black curve, which   is the average of the five validation curves from the cross-validation runs. From the minimum in the black curve,  shown as the vertical gray line, the best epoch is found.}
\label{fig:loss_equally}
\end{figure}

Finally, we show the loss curve in Figure \ref{fig:loss_equally}. The
training losses (dotted lines) and validation losses (solid lines) in different colors correspond to the five different
cross-validation runs. Additionally, we give the
mean of the validation curves with a black solid line. This line is
used to obtain the best epoch, which  in this specific case is epoch 122 (vertical line). The corresponding loss is 0.0528 obtained
with Eq. \ref{eq:loss}.

From the loss curve we see that the network does not overfit much to
the training set since 
the validation curves do not increase much for higher epochs, but still enough to define the optimal epoch to terminate the final training. This is a sign that drop-out is not needed, which is supported by additional tests (see   Sect.~\ref{sec:network}).

\section{Further network tests}
\label{sec:tests}

In addition to the networks described in Sect.~\ref{sec:results}, where we
mainly investigated the effect of the Einstein radius distribution, in this section we discuss
 further tests on the training data set.

\subsection{Data set containing double or quads only}
\label{sec:tests:doubleorquadsonly}

We considered a specialized network for one of the two strong
lensing options and limited our sample to either doubles or quads, where the image multiplicity is based on the centroid of the source (as the spatially extended parts of the source could have different
image multiplicities depending on their positions with respect to the
lensing caustics).
In the case where we limited the sample to doubles only, we  did our standard grid search
for the different hyperparameter combinations for two samples
with naturally distributed Einstein radii above $0.5\arcsec$ and above
$2.0\arcsec$. With these networks we found no notable difference compared
to the sample containing both doubles and quads (see
Sect.~\ref{sec:results:normal0p5} and Sect.~\ref{sec:results:normal2p0}),
which was expected as the doubles   dominate the sample including
both doubles and quads by a factor of around 20-30 (for the different networks depending on the lower limit of the Einstein radii). 

When we limited the sample to quads only, we performed  our grid
search again for the different hyperparameter combinations of  both
samples with naturally distributed Einstein radii above $0.5\arcsec$
and above $2.0\arcsec$ and also with equally distributed Einstein
radii. Since the chance of obtaining four images is smaller than the
chance of observing two images based on the necessary lensing
configuration probability, the sample sizes are smaller with 42 063, 19 176, and 28 398 lensing systems. Therefore,
the output has to be considered with care as this is much lower than
typically used for such a network.

\begin{figure}[ht!]
\includegraphics[trim=20 60 130 1090, clip, width=0.47\textwidth]{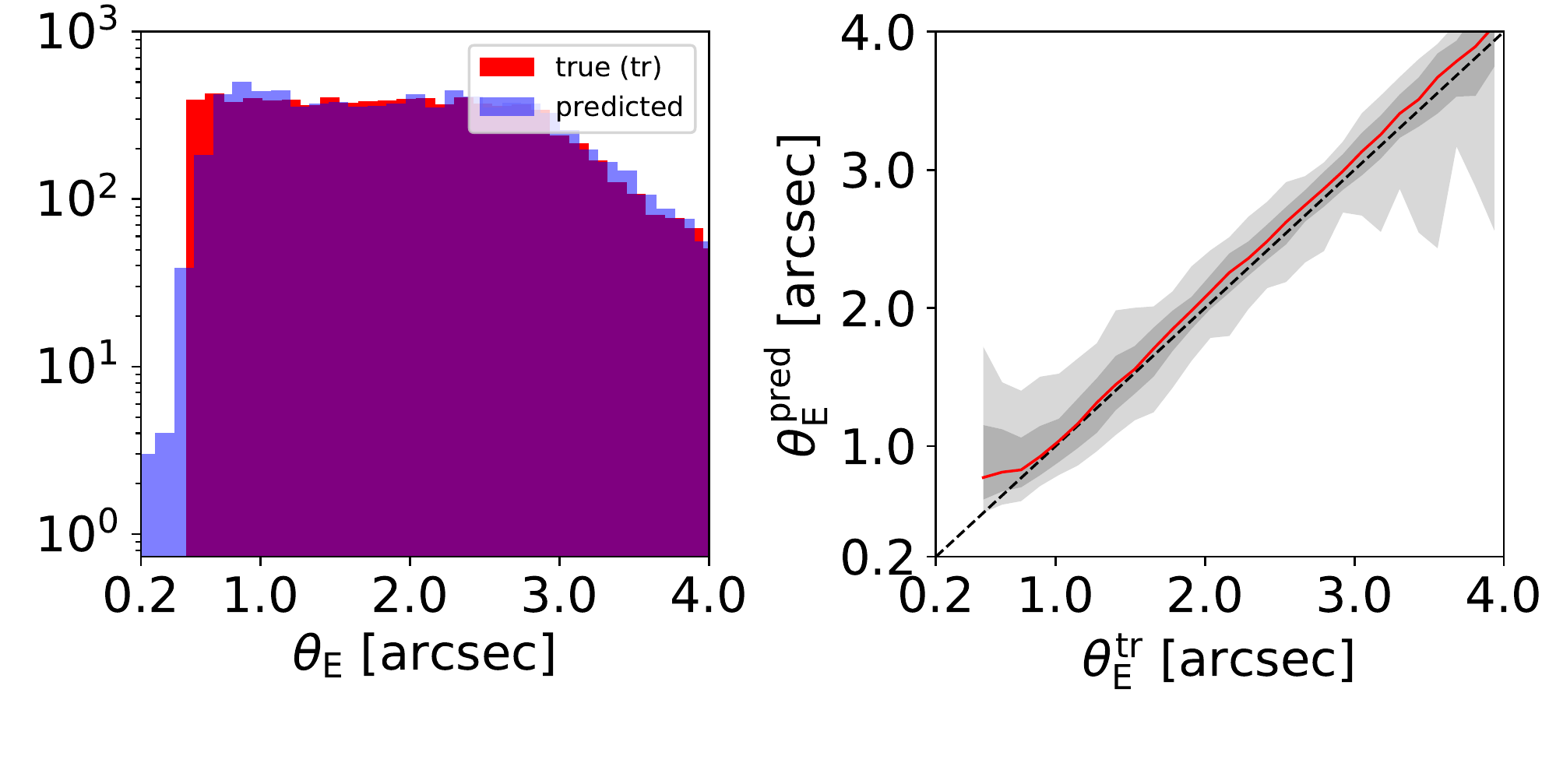}
\caption{Network performance under the assumption of a lowest Einstein
  radius $\theta_\text{E,min}$ of $0.5\arcsec$ but a uniform
  distribution with quadruply lensed images. The left panel shows
  histograms of the ground truth (tr) in red and of the predicted
  values in blue.  The right panel is a 1:1 plot of predicted
  against true Einstein radius. For both plots 30 bins over the plotting range are used.}
\label{fig:comparison_quadsonly}
\end{figure}

It turns out that these networks perform equally well on the lens
center and ellipticity but better for the Einstein radius shown in
Figure \ref{fig:comparison_quadsonly}. By comparing this plot to Figure~\ref{fig:comparison_equally},
  we find the main improvement that the 1$\sigma$ and 2$\sigma$ scatters are substantially reduced and with smaller bias for systems with larger $\theta_{\rm E}$. An improvement on the Einstein radius is expected as the network
gets the same information for the lens, but more for the lensed arcs. Even
if  one image is now too faint to be detected or is too blended with
the lens there are three images from the quad left over to provide
information on the Einstein radius.

To increase the sample we simulated a new quads-only batch with the
source brightness boosted by one magnitude, which resulted in a $\sim
1.5$ times larger sample than before. This is still small compared to
the other double or mixed samples. Now we have a brightness ratio peak
at $\log \left( \frac{I_\text{s}}{I_\text{l} } \right)\sim -1.5$
instead of $\sim -2.0$ (as shown in Figure~\ref{fig:brightness}). The
 performance obtained with this trained network (the loss is 0.0673 for the
network with $w_{\theta_\text{E}}=5$) is still similar to that for the
quads-only network without magnitude boost (the loss is 0.0688) and no
significant performance difference is observed for the individual
parameters.

\subsection{Comparison to lens galaxy images only}
\label{sec:tests:lensonly}

As further proof of the network performance on the Einstein
radius, we test how well the network is able to predict
the parameters from images of only the lens galaxies (i.e., without lensed arcs). As 
expected, the network performs similarly well for the lens center and axis
ratio, but much worse for the Einstein radius with a 1$\sigma$ value of $0.41\arcsec$. This shows us that the
arcs are bright enough and sufficiently deblended from the lens
galaxies to be detectable by the CNN.

\section{Prediction of lensed image position(s) and time-delay(s)}
\label{sec:timedelay}

After obtaining a network for different data sets (see
Table \ref{tab:overview}), we compared the true and predicted parameter
values directly. Since the main advantage of the network is the
computational speed-up compared to recent methods and the fully
automated application, the network is very useful for planning follow-up
observations. This needs to be done   quickly in case there is,
for instance, a SN or a short-lived transient occurring in the background source.  We explore below how accurately we can predict the positions and \mbox{time-delays} of the next appearing SN images.

We used the predicted SIE parameters
from the networks to predict the image positions and \mbox{time-delays} and
compared them to those obtained with the ground-truth SIE model
parameter values. This  gives us a better understanding of how well the 
network performs and if the obtained accuracy is sufficient for such
an application. For this comparison we 
computed the image positions of the true source center based on the true
SIE parameters obtained by the simulation for the sytsems of the test set (hereafter true image
positions). After removing the central highly demagnified lensed image as this would not be observable (given its demagnification and the presence of the
lens galaxy in the optical--infrared), we computed the \mbox{time-delays} for these systems
(hereafter true \mbox{time-delays} $\Delta t^\text{tr}$) by using the known
redshifts and our assumed cosmology. Based on these true image positions and \mbox{time-delays}, we  were able to select the \mbox{first-appearing} image and use its true image position to predict the source position with our predicted SIE mass model.  
 This source position was then used to predict the image positions (hereafter
predicted image positions) of the next-appearing SN images based on the SIE parameter values predicted
with our modeling network. 
The predicted image positions were then used
to predict the \mbox{time-delays} (hereafter predicted \mbox{time-delays} $\Delta t
^\text{pred}$) with the network predicted SIE parameter. We 
directly compared the image positions and \mbox{time-delays} that we obtained with
the true and with the network predicted SIE parameters when we had the
same number of multiple images. If  the number of images did not match, which happened for 7.8\% of the sample used for the network with equally balanced Einstein radii distribution containing double and quads, we omitted the candidate from this analysis as a fair comparison was not possible. Since we always remove the central image, we obtain for a double and quad, respectively, two and four images and one and three
\mbox{time-delays}. Since the \mbox{time-delays} can be very different, we also
compared the fractional difference between the true and predicted \mbox{time-delays} with respect to the true \mbox{time-delays}. 

\begin{figure*}[ht!]
\includegraphics[trim=0 0 0 0, clip, width=0.47\textwidth]{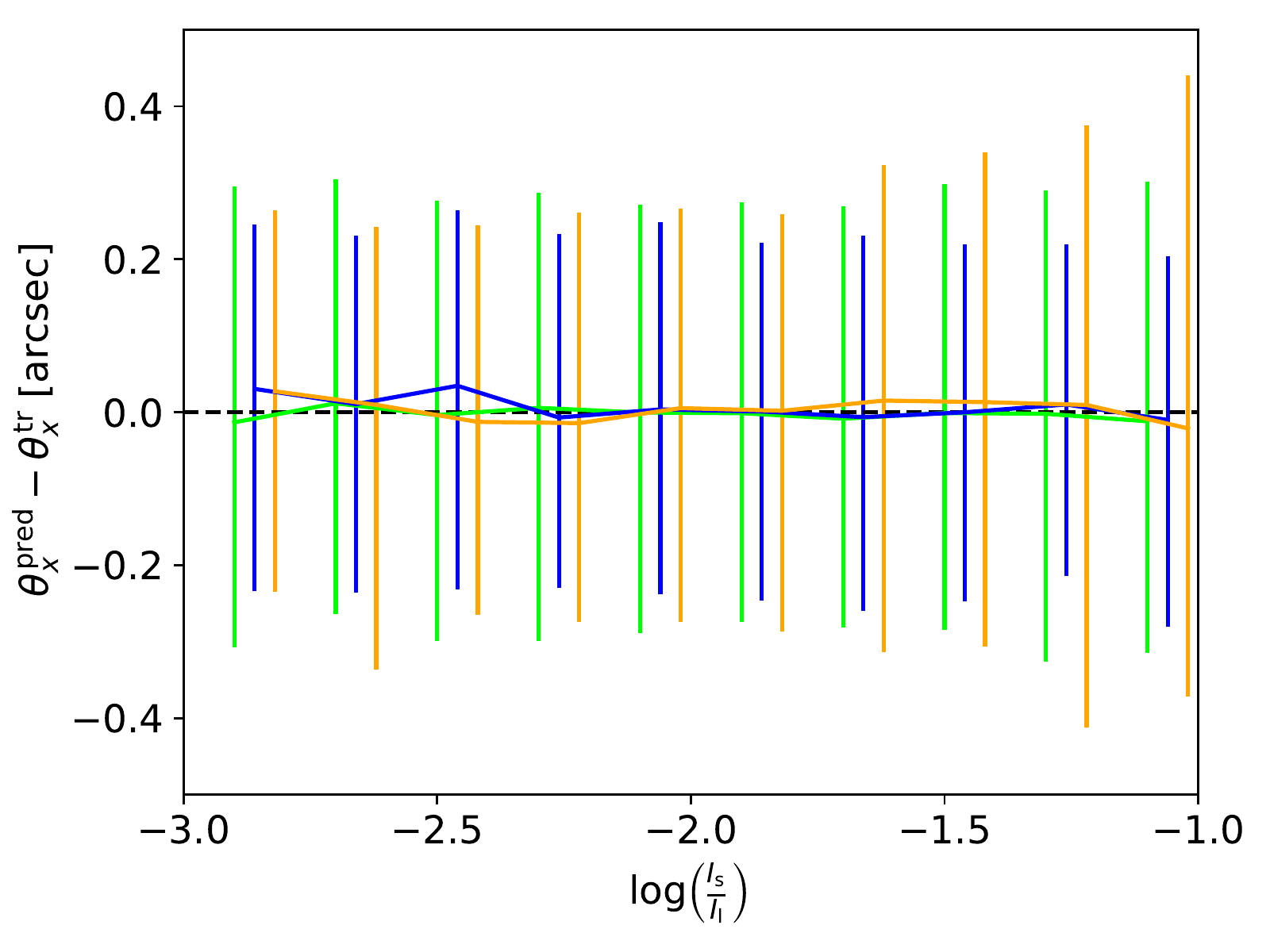}
\includegraphics[trim=0 0 0 0, clip, width=0.47\textwidth]{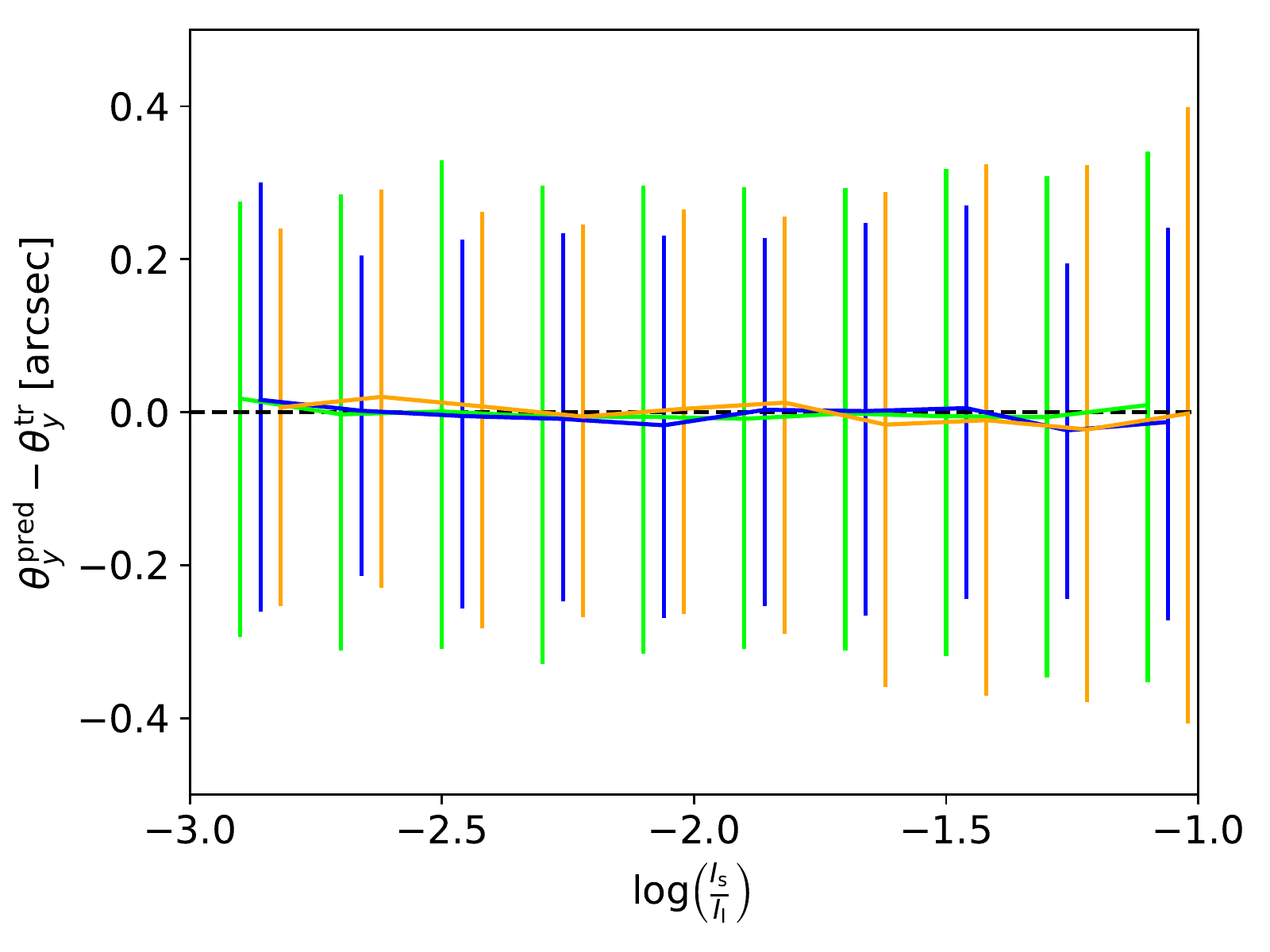}

\includegraphics[trim=0 0 0 0, clip, width=0.47\textwidth]{plots/plot_12063,10166,11084_Legend.pdf}
 \includegraphics[trim=0 0 0 0, clip, width=0.47\textwidth]{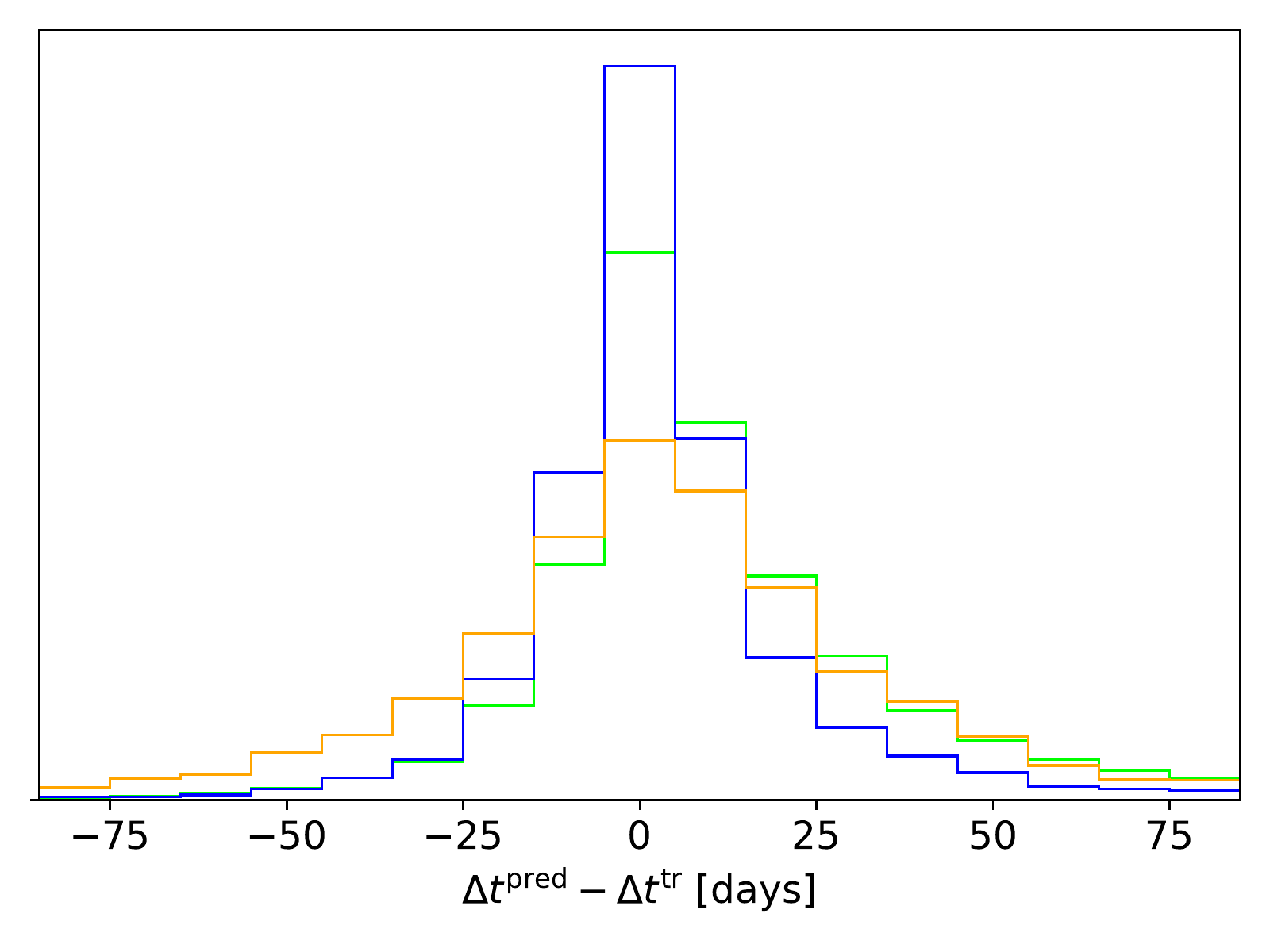}

 \includegraphics[trim=0 0 0 0, clip, width=0.47\textwidth]{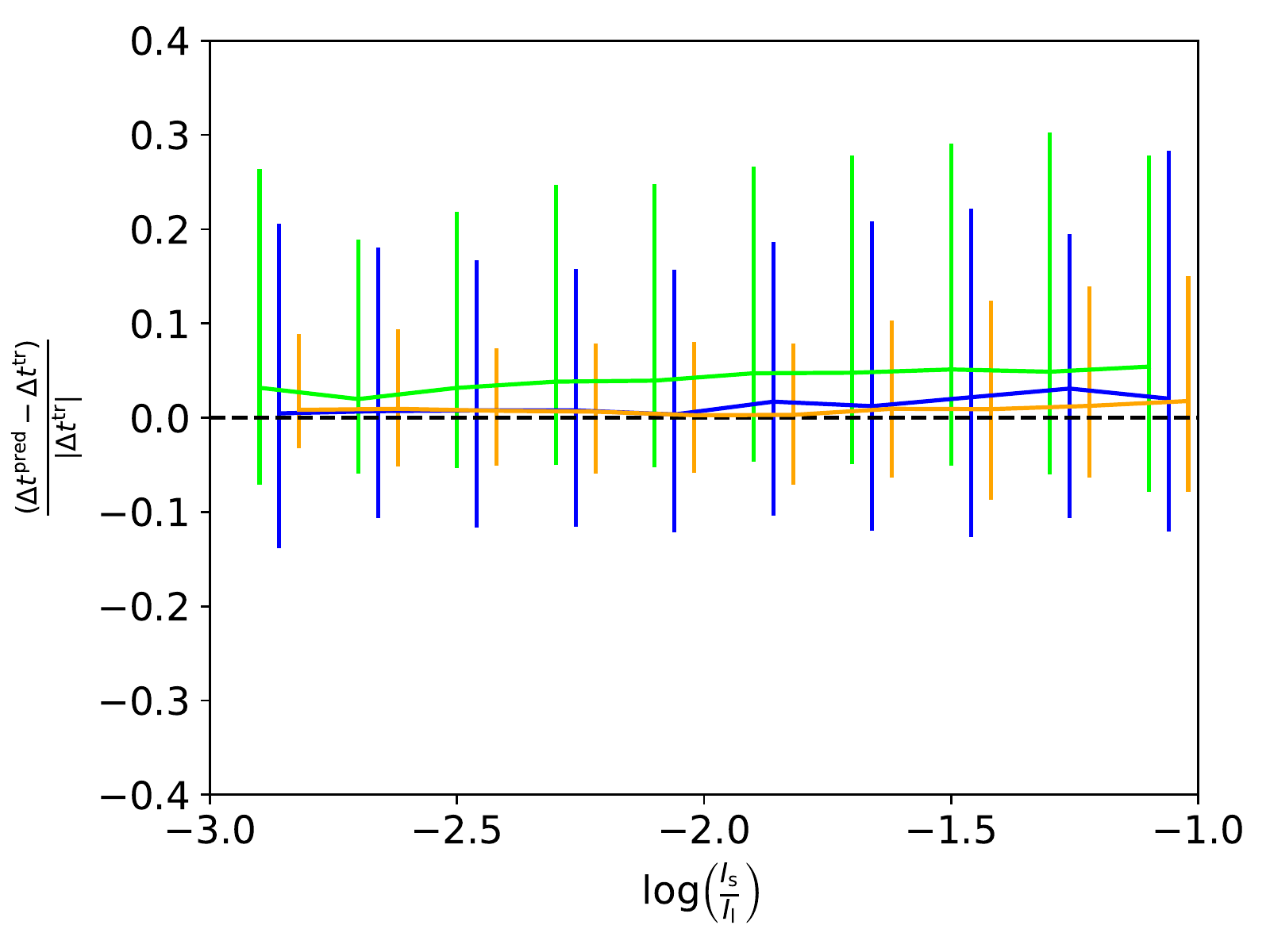}
 \includegraphics[trim=0 0 0 0, clip, width=0.47\textwidth]{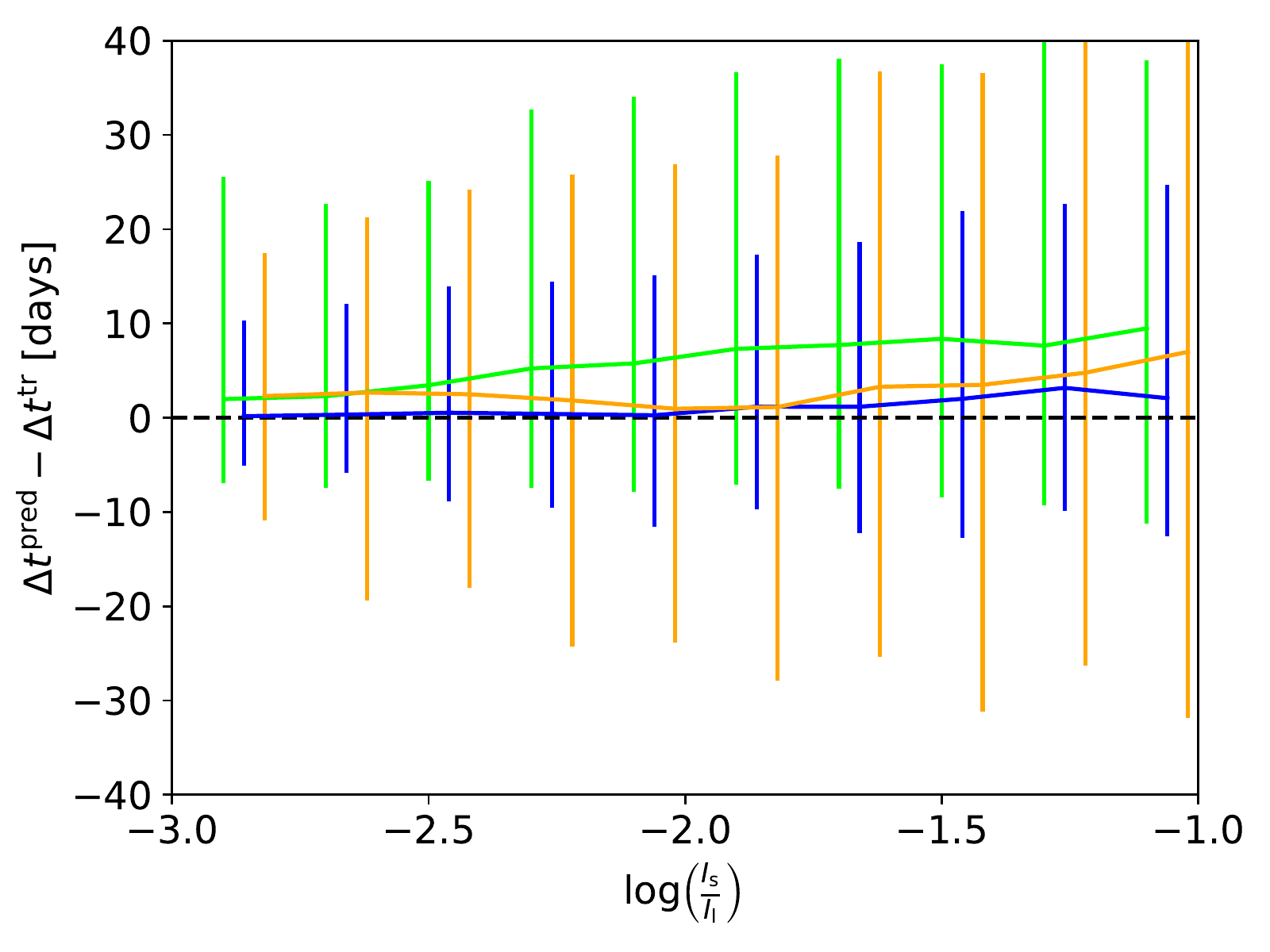}
\caption{Precision of model network predictions as a function of the lens and lensed
  source brightness ratio in the range $-3  \leq \text{log} \left( \frac{I_s}{I_l} \right) \leq -1$ for the three networks presented in Sect.~\ref{sec:results} applied to the restricted sample with $\theta^\text{tr}_\text{E} > 0.8 \arcsec$. The upper row shows the image position offset for the $x$ coordinate (left) and $y$ coordinate (right). In the middle panel is the legend (left) and a histogram of the difference in \mbox{time-delay} (right), while in the bottom row is shown the fraction of the \mbox{time-delay} difference and the true \mbox{time-delay} (left) and the \mbox{time-delay} difference (right). The curves show the median and the vertical bars the 1$\sigma$ values. The blue and orange bars have been shifted slightly to the right for better visualization.}
\label{fig:comparison_im_pos_timedelay}
\end{figure*}

We chose again the three main networks from Sect.~\ref{sec:results} for this
 comparison;   they are shown in Figure \ref{fig:comparison_im_pos_timedelay}. All
 three sets contain quads and doubles, and assume a loss weighting
 factor of 5 for the Einstein radius. The first set assumes a lower
 limit on the Einstein radius of 0.5\arcsec (blue), the second a lower
 limit of 2\arcsec (yellow), and the third a lower limit of again 0.5\arcsec\
 but with a uniform distribution on the Einstein radii instead
 of the natural distribution following the lensing probability
 (green). We plot the quantities as a function of the brightness ratio $\log \left(
   \frac{I_\text{s}}{I_\text{l}} \right)$ in analogy to Figure~\ref{fig:cornerplot} and Figure~\ref{fig:brightness}. 

In detail, Figure \ref{fig:comparison_im_pos_timedelay} contains in the
upper row the median difference in the image position for the $x$
coordinate (left) and $y$ coordinate (right) with the 1$\sigma$ value
per brightness ratio bin, where only the additional image positions
are taken into account as the first reference image is known, and thus they
do not need to be predicted. We obtain for all three networks a median
offset of nearly zero independent of the brightness ratio and whether
we limit further in Einstein radii or not. The 1$\sigma$ values are
around $0.25\arcsec$, corresponding to $\sim 1.5$
pixels. Explicitly, we find for the equally distributed sample applied to $\theta_\text{E}> 0.8\arcsec$ a median image position offset of $(0.00_{-0.29}^{+0.29}) \arcsec$ and $(0.00_{-0.31}^{+0.32}) \arcsec$ for the $x$ and $y$ coordinate, respectively. Interestingly, the 1$\sigma$ values are slightly larger for
quads than doubles as we would have expected that quads provide more
information to constrain the SIE parameter values, and thus predict
 the image positions better. The reason for this is probably because
quads  generally have higher image magnification than doubles, and image
offsets are larger with higher magnification.

\begin{figure*}[ht!]
\includegraphics[trim=0 0 0 0, clip, width=0.47\textwidth]{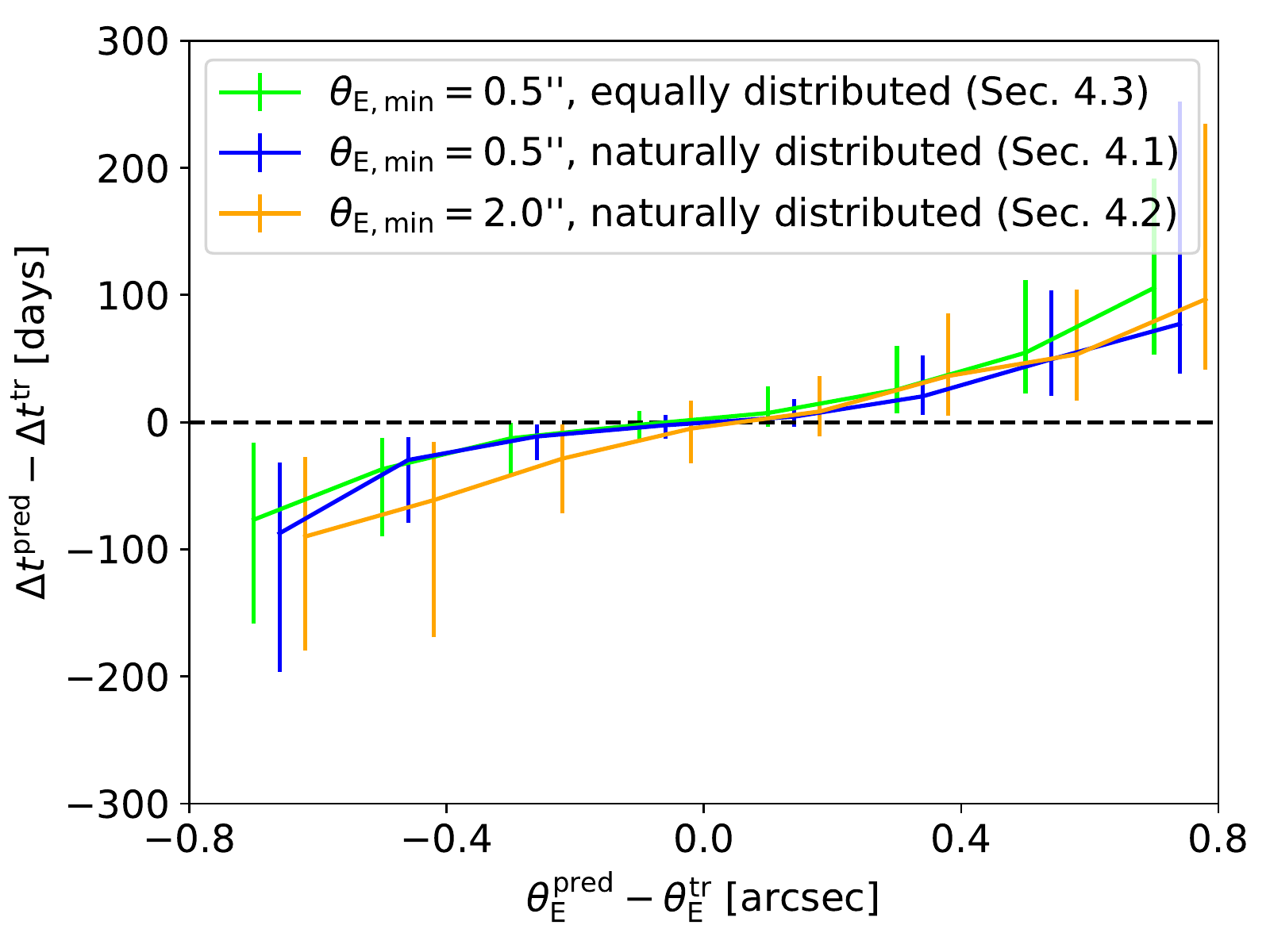}
\includegraphics[trim=0 0 0 0, clip, width=0.47\textwidth]{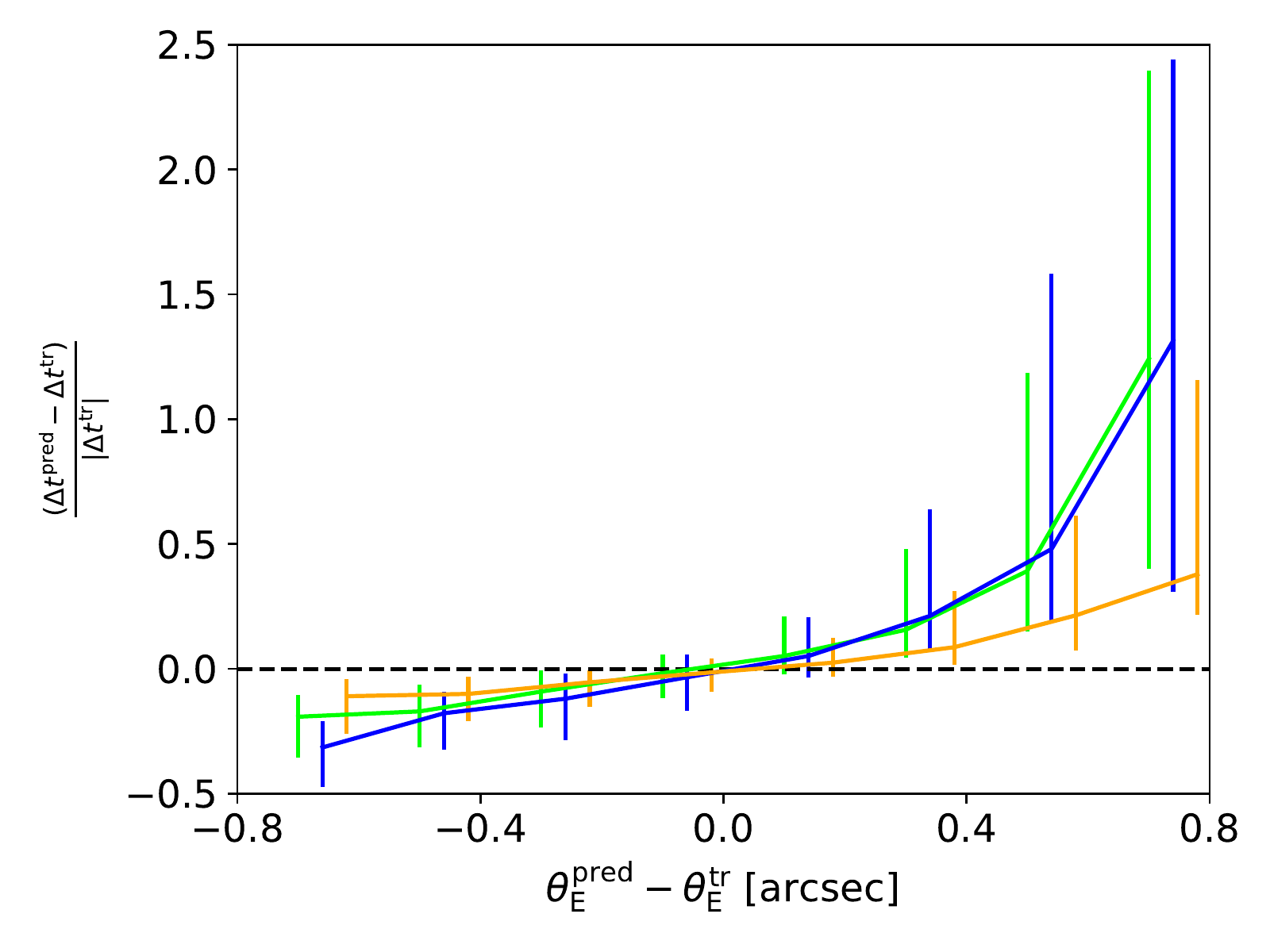}
\caption{Correlation between Einstein radius offset in the range $-0.8\arcsec \leq \theta_\text{E}^\text{pred} - \theta_\text{E}^\text{tr} \leq 0.8\arcsec $ and \mbox{time-delay} difference (left panel) or fractional \mbox{time-delay} difference (right panel) by applying the different networks to their samples after limiting to $\theta_\text{E}^\text{tr}>0.8\arcsec$. The blue and orange bars have
been shifted slightly to the right for better visualization.}
\label{fig:comparison_rE_timedelay}
\end{figure*}

The middle row of Figure~\ref{fig:comparison_im_pos_timedelay} shows
the legend (left) and a histogram of the difference between the
predicted \mbox{time-delay} $\Delta t^\text{pred}$ and the true \mbox{time-delay}
$\Delta t^\text{true}$. The bottom row shows the difference in \mbox{time-delay} divided by the absolute value of the true \mbox{time-delay} per
brightness ratio bin (left) and the difference of the \mbox{time-delays}
again per brightness ratio bin (right). In terms of \mbox{time-delay}
difference, the network trained on the natural distribution (blue)
performs better than that with uniform distribution (green), but
especially for the network trained for lens systems with large
Einstein radius (orange) we obtain notable differences. In
  detail, we obtain a median with 1$\sigma$ value for the naturally
  distributed sample (blue; see Sect. \ref{sec:results:normal0p5}) for the
  \mbox{time-delay} difference of
  $2_{-6}^{+18}$\,days 
  and a fractional \mbox{time-delay} difference 
  of
  $0.05_{-0.09}^{+0.47}$.
  Since we find a strong
  correlation between the offset in the Einstein radius and the \mbox{time-delay} offset (see Figure \ref{fig:comparison_rE_timedelay}),
  we exclude again the very small Einstein radii systems
  ($\theta_\text{E}^\text{tr}< 0.8\arcsec$) and obtain  for the
  \mbox{time-delay} difference
  $1_{-11}^{+18}$\,days 
  and for the fractional difference 
  $0.01_{-0.12}^{+0.19}$. For the equally distributed sample (green; see Sect.~\ref{sec:results:equally}) we obtain, with  $\theta_\text{E} > 0.5\arcsec$ and $\theta_\text{E} > 0.8\arcsec$, respectively, for the \mbox{time-delay} difference 
  $7_{-6}^{+38}$ and $6_{-8}^{+36}$\,days and for the fractional \mbox{time-delay} difference
  $0.06_{-0.05}^{+0.45}$ and $0.04_{-0.05}^{+0.27}$.
  This restriction is easily applicable in practice since  individual lensing systems are
only  followed up   at a given time, and it is possible to check
  by looking at the image of the individual system whether the
  Einstein radius is $>$$0.8\arcsec$. Depending on the predicted \mbox{time-delay}, the model can be further improved by using traditional manual maximum likelihood modeling methods to verify
  the predicted \mbox{time-delay}.

The fractional offset in the predicted \mbox{time-delays} of $0.04_{-0.05}^{+0.27}$
that we achieve with our CNN for systems with $\theta_\text{E} > 0.8\arcsec$ for the uniformly distributed $\theta_{\rm E}$ sample (i.e., with a symmetrized scatter of $\sim$$16\%)$ is close to the limit that would be achievable even
with detailed and \mbox{time-consuming} MCMC models of ground-based images.  This
is because the assumption of the SIE introduces additional
uncertainties on the predicted \mbox{time-delays} in practice, even though detailed
MCMC models of images would typically yield more precise and accurate
estimates for the SIE parameters than our CNN. While galaxy mass
profiles are close to being isothermal, the intrinsic scatter in the
logarithmic radial profile slope $\gamma'$ (where the 3D mass density $\rho(r_{\rm 3D}) \propto r_{\rm
  3D}^{-\gamma'}$) is around $\pm 0.15$, translating to $\sim \, 15\%$
scatter in the \mbox{time-delays} \citep[e.g.,][]{koopmans06, auger10,
  barnabe11}.  In other words, if a lens galaxy has a power-law mass
slope of $\gamma'=2.1$, then our assumed SIE mass profile (with
$\gamma'=2.0$) for it would predict \mbox{time-delays} that are $\sim$$10\%$
too high \citep[e.g.,][]{wucknitz02, suyu12a}.  While constraining the
profile slope $\gamma'$ with better precision than the intrinsic
scatter for individual lenses is possible, this would require
\mbox{high-resolution} imaging from space or ground-based adaptive optics
\citep[e.g.,][]{dye05, chen16}.  Given the difficulties of measuring
the power-law mass slope $\gamma'$ from seeing-limited ground-based
images of lens systems \citep[although see][for the optimistic
  scenario when various inputs are known perfectly such as the point
  spread function]{meng15}, we conclude that our network prediction
for the delays has  uncertainties comparable to that due to the unknown
$\gamma'$.  We expect these two sources of uncertainties to be the
dominant ones in ground-based images.

We also find a decrease in the performance with increase in brightness ratio, which is in the first instance counterintuitive. If we consider the fractional offset in the left panel, we
see a better performance for the sample with an Einstein radius
lower limit of $\theta_\text{E,min} = 2\arcsec$ (orange), especially in terms of the 1$\sigma$ scatter, when compared to the other two networks.  This $\theta_\text{E,min} = 2\arcsec$ network also has minimal bias, as shown by the median line. 
This is understandable
as the \mbox{time-delays} are longer for systems with a bigger Einstein
radius, and therefore the fractional uncertainty is smaller.  The accuracy in \mbox{time-delay} difference (lower right
plot) is good, although the 1$\sigma$ scatter is quite large, $\sim$$20$ days. With
this reasoning, we can also understand the worse performance of the equally
distributed sample (green) compared to the naturally distributed
sample (blue) as it contains a much higher fraction of systems with
bigger image separation. 
As a higher brightness ratio ($\log(I_{\rm s}/I_{\rm l}$)) tends to be associated with systems with higher $\theta_{\rm E}$, the prediction of delays thus has larger scatter, as shown in the bottom right panel.  
Moreover, we 
 note that we find a better performance for doubles than quads,
probably because of smaller image separation and shorter \mbox{time-delays} of quads.

During this evaluation of the networks we should  keep in mind that the
main advantage of these networks is the run time: we need only a few
seconds to estimate
the SIE model parameters, the image positions, and the corresponding \mbox{time-delays}. Therefore, it is expected that we do not
reach the accuracy of current modeling techniques using MCMC sampling
which can take weeks. Nonetheless, the network results can serve as input to conventional modeling and can help speed up the overall modeling.


\section{Comparison to other modeling codes}
\label{sec:otherTechniques}

There are already several modeling codes developed, and they can be separated  into two main groups. The state-of-the-art codes that rely on MCMC sampling have been widely tested and  used for most of the modeling so far. The advantage of such codes are their flexibility in image cutout size or pixel size and also in terms of profiles to describe the lens light or mass distribution. With the advantage of the variety of profiles comes the disadvantage that the codes require a lot of user input which limit the applicability of such codes to a very small sample, or specific lensing systems that are modeled. Moreover, based on the MCMC sampling of the parameter space is very computationally intensive, and thus can take up to weeks per lens system, although some steps can be parallelized and run on multiple cores.

Since the number of known lens systems has grown in the past few years and will increase substantially with upcoming surveys like LSST and Euclid, the codes used to analyze individual lens systems will no longer be sufficient. Thus, the modeling process must be more automated and a speed-up will be necessary. While some newer codes \citep[e.g.,][]{nightingale18, shajib19, ertl20} are automating the modeling steps to minimize the user input, they still rely on sampling the parameter space such that the run time remains on the order of days or weeks, and some user input per lens system. With these codes it   should be possible to  obtain  results that are comparable with the current interactive modeling procedure.

The second new kind of modeling is based on machine learning such as that used in this work. The first network for modeling strong lens systems was presented by \citet{hezaveh17}. While they use Hubble Space Telescope data quality, we use ground-based HSC images with  similar quality to those used by \citet{pearson19} as most of the newly detected lens systems will be in first instance observed with ground-based facilities. Moreover, \citet{hezaveh17} suggest  first removing the lens light and then only modeling  the arcs with the network. Given the differences in the data quality, number of filters, and modeling procedure between \citet{hezaveh17} and our work, we cannot further compare the performance fairly.

\citet{pearson19} consider  modeling with and  without lens light subtraction, but found no notable difference; thus,  we only consider modeling the lens system without an additional step to remove the lens light. Since we provide the image in four different filters, the network is able to distinguish internally between the lens galaxy and the surrounding arcs. In contrast to \citet{pearson19}, we use the SIE
profile with all five different parameters, while they only predict three parameters: axis ratio, position angle (equivalent to our complex ellipticity), and the Einstein radius. Moreover, they completely mock up  their training data, assume a very conservative threshold of S/N$>$20 in at least one band, and do not include neighboring galaxies that can confuse the CNN; instead,  we are more realistic by using real observed images as input for
the simulation pipeline. We further assume an offset between lens mass distribution and lens light distribution, which complicates the CNN parameter inference further as the network has to predict the mass distribution  only from the images. This way we have more realistic lens light and mass
distributions and also include neighboring objects which the network
has to learn to distinguish from the lensing system. \citet{pearson19} make use of a CNN (the same type of network that we use,  and also    \citealt{hezaveh17}),  but they use slightly smaller input cutouts ($57 \times 57$ pixels) and a different network architecture (six convolutional layers and two FC layers) than ours.
They investigated mostly the effect of using a different number of filters and whether to use lens light subtraction, whereas we investigate the effect on the underlying samples and a simulation with real observed images, which means that  we do not have a scenario that assumes the exact same properties. The closest scenario, from \citet{pearson19} the results from LSST-like $gri$ images including lens light and our results based on HSC $griz$ images with naturally distribution of the Einstein radii, shows that both networks are very similar in their overall performance.
The reason that the performance on the Einstein radius by \citet{pearson19} is better and that they do not suffer from the same biases in $\theta_\text{E}^\text{pred}$, even with a non-flat $\theta_\text{E}^\text{tr}$ distribution in their simulations, is perhaps because they use idealized simplistic simulations (high S/N, well-resolved systems, no neighbors). 

There are also other recent publications related to strong lens modeling with machine learning. \citet{bom19}   suggest  a new network that predicts four parameters: the Einstein radius $\theta_\text{E}$, the lens redshift $z_\text{d}$, the source redshift $z_\text{s}$, and the related quantity of the lens velocity dispersion $v_\text{disp}$. They adapt, as we do,  a SIE profile to mock up their training data with an image quality similar to that from the Dark Energy Survey. Comparing their performance on the Einstein radius (Figure~8, left panel, assuming $\theta_\text{E}^\text{tr}$ on the $x$-axis and $\theta_\text{E}^\text{pred}$ on the $y$-axis) to our performance with the natural distribution (Figure~\ref{fig:comparison_normal0p5}), we find a similar trend. Both networks slightly overpredict at the very low end and underpredict at the high end. If we compare this to the network with equal distribution, we see a clear improvement of our network on the median line for $\theta_\text{E} \sim 2-3\arcsec$. Since this code only provides  the Einstein radius instead of a full SIE model, the applicability is somewhat limited.

\citet{madireddy19} suggest a modular network to combine lens detection and lens modeling which to date  have been done  with complete independent networks. In detail, they have four steps. The first  is to reduce the background noise (so-called image denoising), followed by a lens light subtracting step (the  deblending step), before the next network decides whether this is a lens system or not. If it detects the input image to be a lens, the module is called to predict the mass model parameter values. Each module of the network is a very deep network and both modules for detection and modeling make use of the residual neural network (ResNet) approach. They use a sample of 120,000 images, with 60,000 lenses and 60,000 non-lenses, and split this into 90\% and 10\% for the training and test set, respectively, without making use of the cross-validation procedure. \citet{madireddy19} use, as do  \citet{pearson19}, completely mocked-up images based on a SIE profile with fixed centroid to the image center such that the modeling module predicts three quantities, Einstein radius, and the two components of the complex ellipticity. Based on the different assumptions a direct comparison of the lens modeling performance is not possible. However, we see that the performance is typical for the current state of CNNs based on \citet{pearson19}.

Comparing the network-based modeling to the traditional model using MCMC on a concrete sample is difficult as   first we have to obtain the mass models for that sample with both methods. However, in general it is expected that the MCMC models are typically more precise than those obtained with neural networks because of the interactive and individual modeling procedure. In the MCMC modeling,   the image residuals can be inspected to see whether the model is good and trustworthy, or if the parameters  need to be refined  further and  different mass profile adopted (e.g., SIE plus external shear). In contrast, the fully automated procedure with CNN does not inspect the individual images and residuals in detail. However, for upcoming surveys like LSST it is impossible to model all the expected $\sim$$100,000$ lens systems in the traditional MCMC way systematically given the computational time required.  The only way to analyze the entire LSST lens sample will be a fully automated, fast procedure where a small fraction of outliers and (probably) slightly higher uncertainties are acceptable.  Therefore, the substantial speed-up is a very important advantage of CNN modeling, as we can process one lens systems with our CNN in fractions of a second compared to weeks  or months with MCMC methods.  If one is interested in a specific lens system such as a lensed SN, one can consider starting with a CNN to get a good initial mass model and then refining with traditional methods to achieve a good balance between speed and accuracy.

\section{Summary and conclusion}
\label{sec:conclusion}

In this paper, we presented a convolutional neural network to model
in a fully automated way and very quickly the mass distribution of galaxy-scale
strong lens systems by assuming a SIE profile. The network is trained
on images of lens systems generated with our newly developed code that takes real
observed galaxy images as input for the source galaxy (in our case from the Hubble Ultra
Deep Field), lenses the source onto the lens plane, and adds it to
another real observed galaxy image for the lens galaxy (in our case from the HSC
SSP survey). We chose the HSC images as lenses and adopted their pixel sizes
of $0.168\arcsec$ as this is similar to the data
quality expected from LSST. With this procedure we simulated different samples
to train our networks where we distinguish between the lens types
(quads+doubles, doubles-only, and quads-only) and on the lower
limit of the Einstein radius range. Since we find a strong dependence on the
Einstein radius distribution, we also consider a uniformly distributed
sample and also a weighting factor of 5 for the Einstein radius'
contribution to the loss. With this we obtain eight
different samples for each of the two different weighting assumptions
summarized in Table \ref{tab:overview}.

For each sample we then
perform a grid search to test different hyperparameter combinations to obtain the best
network for each sample, although we find that the CNN performance depends much more critically on the assumptions of the mock training data (e.g., quads, doubles, both, or Einstein radius distribution) rather than on the fine-tuning of hyperparameters. From the different networks presented in Table \ref{tab:overview}, we find a good improvement for the networks trained with quads-only compared to the networks trained on both quads and doubles. If the system type is known, we therefore recommend using the corresponding network. Since the Einstein radius is a key parameter, we weighted its loss higher than for the others and, although the minimal validation loss is higher, we advocate these networks for modeling HSC-like lenses. With the network trained on both quads and doubles with the uniform distributed sample of $\theta_\text{E}$ $>0.5\arcsec$, we obtain for the five SIE parameters a median with 1$\sigma$ value as follows:
      $\Delta x = (0.00^{+0.30}_{-0.30}) \arcsec$,
      $\Delta y = (0.00^{+0.30}_{-0.29} ) \arcsec $,
      $\Delta \theta_\text{E} = (0.07^{+0.29}_{-0.12}) \arcsec$,
      $\Delta e_x = -0.01^{+0.08}_{-0.09}$, and
      $\Delta e_y = 0.00^{+0.08}_{-0.09}$.

After comparing the network performance on the SIE parameter level, we
tested the network performance on the lensed image and \mbox{time-delay} level. For
this we used the first appearing image of the true mass model to predict
the source position based on the predicted SIE parameter. From this source position and the network
predicted SIE parameters, we then predicted the other image position(s) and \mbox{time-delay(s)}. We find for the sample with doubles and quads a uniform
distribution in Einstein radii and a weighting factor
$w_{\theta_\text{E}}$ of five by applying the network to $\theta_\text{E}>0.8\arcsec$ an average image offset of $\Delta \theta_x =(0.00_{-0.29}^{+0.29})  \arcsec $ and $\Delta \theta_y = (0.00_{-0.31}^{+0.32}) \arcsec $, while we achieve the fractional \mbox{time-delay} difference of
$0.04_{-0.05}^{+0.27}$. 

This is very good given that
we use a simple SIE profile and need only a few seconds per lens
system in comparison to  \mbox{state-of-the-art} methods that require at least days and some user input per lens system. 
We anticipate that fast CNN modeling such as the one developed here will be crucial for coping with the vast amount of data from upcoming imaging surveys. For future work, we suggest  investigating further into creating even more realistic training data (e.g., allowing for an external shear component in the lens mass model) and also  exploring the effect of deeper or more complex network architectures. The outputs of even the network presented here can be used to prune down the sample for specific scientific studies, which can then be followed up with more detailed conventional mass modeling techniques.

\FloatBarrier
\begin{acknowledgements}
  We thank Y.~D.~Hezaveh, L.~Perreault~Levasseur, J.~H.~H.~Chan and D.~Sluse for good discussions.
SS, SHS, and RC thank the Max Planck Society for support through the
Max Planck Research Group for SHS. This project has received funding
from the European Research Council (ERC) under the European Unions
Horizon 2020 research and innovation programme (LENSNOVA: grant
agreement No 771776).
This research is supported in part by the Excellence Cluster ORIGINS which is funded by the Deutsche Forschungsgemeinschaft (DFG, German Research Foundation) under Germany's Excellence Strategy -- EXC-2094 -- 390783311.
\\
Based on observations made with the NASA/ESA Hubble Space Telescope, obtained from the data archive at the Space Telescope Science Institute. STScI is operated by the Association of Universities for Research in Astronomy, Inc. under NASA contract NAS 5-26555\\
The Hyper Suprime-Cam (HSC) collaboration includes the astronomical communities of Japan and Taiwan, and Princeton University. The HSC instrumentation and software were developed by the National Astronomical Observatory of Japan (NAOJ), the Kavli Institute for the Physics and Mathematics of the Universe (Kavli IPMU), the University of Tokyo, the High Energy Accelerator Research Organization (KEK), the Academia Sinica Institute for Astronomy and Astrophysics in Taiwan (ASIAA), and Princeton University. Funding was contributed by the FIRST program from Japanese Cabinet Office, the Ministry of Education, Culture, Sports, Science and Technology (MEXT), the Japan Society for the Promotion of Science (JSPS), Japan Science and Technology Agency (JST), the Toray Science Foundation, NAOJ, Kavli IPMU, KEK, ASIAA, and Princeton University. This paper makes use of software developed for the Rubin Observatory Legacy Survey in Space and Time (LSST). We thank the LSST Project for making their code available as free software at http://dm.lsst.org This paper is based in part on data collected at the Subaru Telescope and retrieved from the HSC data archive system, which is operated by Subaru Telescope and Astronomy Data Center (ADC) at National Astronomical Observatory of Japan. Data analysis was in part carried out with the cooperation of Center for Computational Astrophysics (CfCA), National Astronomical Observatory of Japan. We make partly use of the data collected at the Subaru Telescope and retrieved from the HSC data archive system, which is operated by Subaru Telescope and Astronomy Data Center at National Astronomical Observatory of Japan.\\
Funding for the Sloan Digital Sky Survey IV has been provided by the Alfred P. Sloan Foundation, the U.S. Department of Energy Office of Science, and the Participating Institutions. SDSS-IV acknowledges support and resources from the Center for High-Performance Computing at the University of Utah. The SDSS web site is www.sdss.org.\\
SDSS-IV is managed by the Astrophysical Research Consortium for the
Participating Institutions of the SDSS Collaboration including the
Brazilian Participation Group, the Carnegie Institution for Science,
Carnegie Mellon University, the Chilean Participation Group, the
French Participation Group, Harvard-Smithsonian Center for
Astrophysics, Instituto de Astrof\'isica de Canarias, The Johns
Hopkins University, Kavli Institute for the Physics and Mathematics of
the Universe (IPMU) / University of Tokyo, the Korean Participation
Group, Lawrence Berkeley National Laboratory, Leibniz Institut f\"ur
Astrophysik Potsdam (AIP), Max-Planck-Institut f\"ur Astronomie (MPIA
Heidelberg), Max-Planck-Institut f\"ur Astrophysik (MPA Garching),
Max-Planck-Institut f\"ur Extraterrestrische Physik (MPE), National
Astronomical Observatories of China, New Mexico State University, New
York University, University of Notre Dame, Observat\'ario Nacional /
MCTI, The Ohio State University, Pennsylvania State University,
Shanghai Astronomical Observatory, United Kingdom Participation Group,
Universidad Nacional Aut\'onoma de M\'exico, University of Arizona,
University of Colorado Boulder, University of Oxford, University of
Portsmouth, University of Utah, University of Virginia, University of
Washington, University of Wisconsin, Vanderbilt University, and Yale
University.
\end{acknowledgements}

\bibliographystyle{aa}
\bibliography{NetModel}


\end{document}